\documentclass[11pt, oneside]{article}

\usepackage{latexsym}
\usepackage{amssymb,amsmath,amsthm,amsfonts}
\usepackage{mathrsfs}
\usepackage{graphicx}
\usepackage[latin1]{inputenc}
\usepackage{chicago}

\usepackage{color}
\usepackage{rotating}
\usepackage[table]{xcolor}
\usepackage{arydshln}

\theoremstyle{definition}
\newtheorem{Theorem}{Theorem}[section]
\newtheorem{Example}[Theorem]{Example}
\newtheorem{Definition}[Theorem]{Definition}

\newtheorem{Remark}[Theorem]{Remark}

\numberwithin{equation}{section}

\newcolumntype{C}[1]{>{\centering\arraybackslash}p{#1}}

\begin{document}

\title{COPAR -- Multivariate time series modeling using\\ the COPula AutoRegressive model}
\author{Eike Christian Brechmann\footnote{Center for Mathematical Sciences, Technische Universit\"at M\"unchen,
Boltzmannstr. 3, D-85747 Garching, Germany.} \footnote{\textit{Corresponding author.} E-mail: brechmann@ma.tum.de. Phone: +49 89 289-17425.}, Claudia Czado\footnotemark[1] }
\maketitle

\begin{abstract}

Analysis of multivariate time series is a common problem in areas like finance and economics.
The classical tool for this purpose are vector autoregressive models.
These however are limited to the modeling of linear and symmetric dependence.
We propose a novel copula-based model which allows for non-linear and asymmetric modeling of serial as well as between-series dependencies.
The model exploits the flexibility of vine copulas which are built up by bivariate copulas only.
We describe statistical inference techniques for the new model and demonstrate its usefulness in three relevant applications:
We analyze time series of macroeconomic indicators, of electricity load demands and of bond portfolio returns.

\end{abstract}

\section{Introduction}

The analysis of multiple time series is of fundamental interest in finance and economics.
Classically, interdependencies among multivariate time series have been modeled using vector autoregressive (VAR) models. 
Such models provide insights into the dynamic relationship of the time series and often produce forecasts superior to independent univariate models.
VAR models in economics were advocated by \citeN{Sims1980}, standard reference books are \citeN{Luetkepohl2005}, \citeN{Hamilton1994} and \citeN{Tsay2002}.

The bivariate $p$th order vector autoregressive model, VAR($p$), for two time series $\{X_t\}$ and $\{Y_t\}$ is defined as  
\begin{equation}
\begin{pmatrix}X_t\\Y_t\end{pmatrix} = \begin{pmatrix}c_1\\c_2\end{pmatrix} + \Phi_1 \begin{pmatrix}X_{t-1}\\Y_{t-1}\end{pmatrix} + ... + \Phi_p \begin{pmatrix}X_{t-p}\\Y_{t-p}\end{pmatrix} + \begin{pmatrix}\varepsilon_1\\ \varepsilon_2\end{pmatrix},
\label{eq:var}
\end{equation}
where $\Phi_j,\ j=1,...,p,$ are 2-by-2-matrices of autoregressive coefficients and $c_1$ and $c_2$ are constants.
The vector $\boldsymbol{\varepsilon}_t=(\varepsilon_1,\varepsilon_2)^\prime$ is multivariate white noise, that is $E(\boldsymbol{\varepsilon}_t)=\boldsymbol{0}$ and $E(\boldsymbol{\varepsilon}_t\boldsymbol{\varepsilon}_s) = \Sigma$ for $t=s$ and 0 otherwise, where $\Sigma$ is a symmetric positive definite 2-by-2-matrix.
Typically $\boldsymbol{\varepsilon}_t\sim N_2(\boldsymbol{0},\Sigma)$ is assumed.

While VAR models can only capture linear and symmetric dependence in time and between series, we propose a new copula-based model which overcomes such limitations and allows for an extremely flexible modeling.
Copulas are the canonical statistical tool for statistical dependence modeling.
The theorem by \citeN{Sklar1959} shows that every multivariate distribution can be represented in terms of a copula which couples the univariate marginal distributions.
For a random vector $\mathbf{X}=(X_1,...,X_d)\sim F$ with marginal distributions $F_i,\ i=1,...,d$, it is 
\begin{equation}
F(x_1,...,x_d)=C(F_1(x_1),...,F_d(x_d)),
\label{eq:sklar}
\end{equation}
where $C$ is some appropriate $d$-dimensional copula, a multivariate distribution on the unit hypercube with uniform margins (see \citeN{Nelsen2006} and \citeN{Joe1997} for more details).

We will use a fully integrated copula model to capture effects in time and between series.
In particular, our model is built upon a so-called vine copula (see \citeN{KurowickaJoe2011} for an overview).
Such vine copulas are flexible multivariate copulas constructed through a sequence of bivariate copulas, a pair-copula decomposition.
While \shortciteN{SmithMinCzadoAlmeida2010} recently showed how univariate time series can be modeled using a so-called D-vine pair-copula decomposition, we will show how such pair-copula decompositions can be conveniently used to model the dependence among multiple time series.

The contributions of this paper are as follows:
We introduce the so-called copula autoregressive model, COPAR, which exploits the enormous flexibility of vine copula models and allows for non-linear and non-symmetric modeling of serial and between-series dependence.
By allowing for arbitrary marginal distributions, the model can also account for common features of univariate economic and financial time series like skewness and heavy-tailedness which are not captured appropriately using a normal distribution.
Required statistical inference techniques for the model are presented and described in detail.
In addition, we also discuss how the model can be easily used to test for Granger causality, a central concept to determine interdependencies among multiple time series.
The usefulness of our model is demonstrated and carefully evaluated in three relevant applications:
We analyze monthly macro-economic indicators, daily electricity load demands as well as monthly bond portfolio returns.

The paper is structured as follows.
In Section \ref{sect:techbg} we establish the relevant technical background on copulas and pair-copula constructions in particular.
The copula autoregressive model is introduced and discussed in detail in Section \ref{sect:copar}.
The three applications are subsequently treated in Section \ref{sect:applic}, while Section \ref{sect:concl} concludes with an outlook to future research.


\section{Pair-copula decompositions for univariate time series}\label{sect:techbg}

Let $\{X_t\}_{t=1,...,T}$ be a univariate time series of continuously distributed data.
Through conditioning the joint distribution of $\{X_t\}$ can be decomposed as
\begin{equation}
f(x_1,...,x_T) = f(x_1)\prod_{t=2}^T f_{t|1:(t-1)}(x_t|x_1,...,x_{t-1}),
\label{eq:dvinecond}
\end{equation}
where $f$ denotes the common marginal density of $X_t,\ t=1,...,T$; $F$ will denote the corresponding distribution function.
Here we use $r:s := (r,r+1,...,s-1,s)$ for $r<s$ and $f_{s|D}$ denotes the conditional density of $X_s$ given $\{X_r, r\in D\}$.

As outlined in \shortciteN{SmithMinCzadoAlmeida2010} this expression can be used to obtain a general decomposition in terms of bivariate copulas.
For the distribution of $X_s$ and $X_t,\ s<t,$ given $\{X_{s+1},...,X_{t-1}\}$ (to shorten notation we often write $X_{(s+1):(t-1)}$) it follows according to Sklar's theorem \eqref{eq:sklar} that 
\begin{align*}
f_{s,t|(s+1):(t-1)}&(x_s,x_t|x_{s+1},...x_{t-1}) =\\
& c_{s,t|(s+1):(t-1)}(F_{s|(s+1):(t-1)}(x_s|x_{s+1},...,x_{t-1}),F_{t|(s+1):(t-1)}(x_t|x_{s+1},...,x_{t-1}))\\
& \times f_{s|(s+1):(t-1)}(x_s|x_{s+1},...,x_{t-1}) \times f_{t|(s+1):(t-1)}(x_t|x_{s+1},...,x_{t-1}),
\end{align*}
where $c_{s,t|(s+1):(t-1)}$ is an appropriate bivariate copula density.
Rearranging terms gives
\begin{align*}
f_{t|s:(t-1)}&(x_t|x_s,...x_{t-1}) =\\
& c_{s,t|(s+1):(t-1)}(F_{s|(s+1):(t-1)}(x_s|x_{s+1},...,x_{t-1}),F_{t|(s+1):(t-1)}(x_t|x_{s+1},...,x_{t-1}))\\
& \times f_{t|(s+1):(t-1)}(x_t|x_{s+1},...,x_{t-1}).
\end{align*}
Let $u_{r|(r+1):(t-1)}:=F_{r|(r+1):(t-1)}(x_r|x_{r+1},...,x_{t-1})$ and $u_{t|(r+1):(t-1)}:=F_{t|(r+1):(t-1)}(x_t|x_{r+1},...,x_{t-1})$.
By recursive conditioning on $s=1,2,...,t-1$ one then obtains
\begin{equation}
\begin{split}
f_{t|1:(t-1)}&(x_t|x_1,...x_{t-1}) =\\
& f(x_t)c_{t-1,t}(F(x_{t-1}),F(x_t))\prod_{r=1}^{t-2}  c_{r,t|(r+1):(t-1)}(u_{r|(r+1):(t-1)},u_{t|(r+1):(t-1)}).
\end{split}
\label{eq:dvinerec}
\end{equation}
Finally, plugging Equation \eqref{eq:dvinerec} into Expression \eqref{eq:dvinecond} gives
\begin{equation*}
f(x_1,...,x_T) = \prod_{t=1}^T f(x_t)\prod_{t=2}^T \prod_{r=1}^{t-1}  c_{r,t|(r+1):(t-1)}(u_{r|(r+1):(t-1)},u_{t|(r+1):(t-1)}).
\end{equation*}
This is the product of $T(T-1)/2$ bivariate copulas, so-called pair-copulas, and the marginal densities evaluated at each time point $x_t,\ t=1,...,T$.
The construction does not require the selection of a particular copula family, so that very flexible models can be deduced from it.
However, it is clear that copulas corresponding to the same time lag have to be identical.
For example $C_{t-2,t-1}$ and $C_{t-1,t}$ must not be different.

The above construction is called pair-copula decomposition and was introduced by \shortciteN{AasCzadoFrigessiBakken2009}.
The particular way described here is called D-vine and belongs to the more general class of regular vines (R-vines) introduced by \citeN{Joe1996} and \citeN[2002]{BedfordCooke2001}\nocite{BedfordCooke2002} and described in more detail in \citeN{KurowickaCooke2006} and \citeN{KurowickaJoe2011}.
R-vines are a graphic theoretic model to determine which pairs are included in a pair-copula decomposition.
The following definition is taken from \citeN{KurowickaCooke2006}.

\begin{Definition}[Regular vine]

A regular vine (R-vine) on $d$ variables is a sequence of linked trees (connected acyclic graphs) $\mathcal{T}_1,...,\mathcal{T}_{d-1}$ with nodes $\mathcal{N}_i$ and edges $\mathcal{E}_i$ for $i=1,...,d-1$ which satisfy the following three conditions.
\begin{enumerate}

\item Tree $\mathcal{T}_1$ has nodes $\mathcal{N}_1=\{1,...,d\}$.

\item For $i=2,..., d-1$ tree $\mathcal{T}_i$ has nodes $\mathcal{N}_i=\mathcal{E}_{i-1}$.

\item If two edges in tree $\mathcal{T}_i$ are joined in tree $\mathcal{T}_{i+1}$, they must share a common node in tree $\mathcal{T}_i$.

\end{enumerate}
The last property is called proximity condition.

\end{Definition}

\begin{figure}
\centering
\includegraphics[width=.7\linewidth]{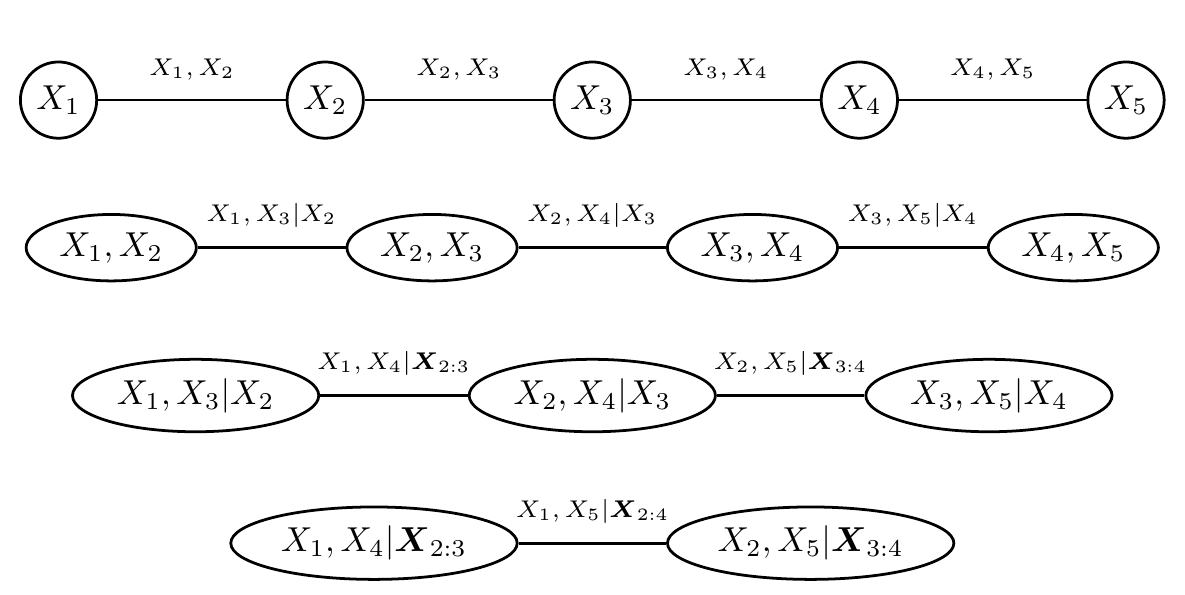}
\caption{D-vine for $T=5$ with edge labels.}
\label{fig:dvine5}
\end{figure}

D-vines are R-vines, where each node is connected to at most two other nodes.
The D-vine corresponding to the above decomposition is shown in Figure \ref{fig:dvine5}.

By associating each edge $e=j(e),k(e)|D(e)$ in an R-vine with a bivariate copula density $c_{j(e),k(e)|D(e)}$, the complete pair-copula decomposition is defined.
The nodes $j(e)$ and $k(e)$ are called conditioned nodes and $D(e)$ the conditioning set, where in each tree from top to bottom an additional variable is added in the conditioning set of the bivariate copula.

\begin{Theorem}[R-vine density (Kurowicka and Cooke 2006, Theorem 4.2)]

The joint density of $X_1,...,X_d$ is uniquely determined and given by
\begin{equation}
f(x_1,...,x_d)= \left[\prod_{k=1}^d f_k(x_k)\right] \times 
\left[\prod_{i=1}^{d-1} \prod_{e \in \mathcal{E}_i} c_{j(e),k(e)|D(e)}(F(x_{j(e)}|\boldsymbol{x}_{D(e)}),F(x_{k(e)}|\boldsymbol{x}_{D(e)}))\right],
\label{eq:rvinedens}
\end{equation}
where $\boldsymbol{x}_{D(e)}$ denotes the sub-vector of $\boldsymbol{x}=(x_1,...,x_d)^\prime$ determined by the indices in $D(e)$.

\end{Theorem}

In \eqref{eq:rvinedens} the arguments of copulas in tree $\mathcal{T}_i$ can be recursively computed from copulas in trees $\mathcal{T}_1,...,\mathcal{T}_{i-1}$ using the general formula
\begin{equation}
F(x|\boldsymbol{v})=\frac{\partial C_{xv_j|\boldsymbol{v}_{-j}}(F(x|\boldsymbol{v}_{-j}),F(v_j|\boldsymbol{v}_{-j}))}{\partial F(v_j|\boldsymbol{v}_{-j})},
\label{eq:hfunc}
\end{equation}
where $C_{xv_j|\boldsymbol{v}_{-j}}$ is a bivariate copula, $v_j$ is an arbitrary component of $\boldsymbol{v}$ and $\boldsymbol{v}_{-j}$ denotes the vector $\boldsymbol{v}$ excluding $v_j$.

To facilitate statistical inference of R-vines, they can be conveniently stored in matrix notation as recently proposed by \citeN{MoralesBook2011} and further explored by \shortciteN{DissmannBrechmannCzadoKurowicka2012}.
Let $M\in\{0,...,d\}^{d\times d}$ be a lower triangular matrix, where the diagonal entries of $M$ are the numbers $1,...,d$ in decreasing order.
In this matrix, according to technical conditions, each row from the bottom up represents a tree, where the conditioned set is identified by a diagonal entry and by the corresponding column entry of the row under consideration, while the 
conditioning set is given by the column entries below this row.
Corresponding copula types and parameters can conveniently be stored in matrices related to $M$. 
The fixed ordering of diagonal entries ensures uniqueness of the R-vine matrix.

The serial D-vine decomposition presented above can be stored in the following matrix
\begin{equation*}
  \left(
  \begin{array}{C{0.9cm}C{0.9cm}C{0.9cm}C{0.9cm}C{0.9cm}C{0.9cm}C{0.9cm}}
		\cellcolor{gray!10}$X_T$\\
		$X_1$ & \cellcolor{gray!10}$X_{T-1}$\\
		$X_2$ & $X_1$ & $\ddots$\\
		$\vdots$ & $\vdots$ & $\ddots$ & $\ddots$\\
		$\vdots$ & $\vdots$ & & $\ddots$ & \cellcolor{gray!10}$X_3$\\
		$X_{T-2}$ & $X_{T-3}$ & $\cdots$ & $\cdots$ & $X_1$ & \cellcolor{gray!10}$X_2$\\
		$X_{T-1}$ & $X_{T-2}$ & $\cdots$ & $\cdots$ & $X_2$ & $X_1$ & \cellcolor{gray!10}$X_1$\\
  \end{array}
  \right),
\end{equation*}
which is easily extendible to include future observations $\{X_{T+1},X_{T+2},...\}$.
For example, the second entry in the first column identifies the conditioned pair $X_1$ and $X_T$ given $\{X_2,...,X_{T-1}\}$.

Corresponding copula types are stored in the off-diagonal entry associated with the pair:
\begin{equation*}
  \left(
  \begin{array}{C{1.8cm}C{2.2cm}C{1.2cm}C{1.2cm}C{1.2cm}C{1.2cm}C{1.2cm}}
		\\
		$C_{1,T|2:(T-1)}$ \\
		$C_{2,T|3:(T-1)}$ & $C_{1,T-1|2:(T-2)}$ \\
		$\vdots$ & $\vdots$ & $\ddots$ \\
		$\vdots$ & $\vdots$ & & $\ddots$ \\
		$C_{T-2,T|T-1}$ & $C_{T-3,T-1|T-2}$ & $\cdots$ & $\cdots$ & $C_{1,3|2}$ \\
		$C_{T-1,T}$ & $C_{T-2,T-1}$ & $\cdots$ & $\cdots$ & $C_{2,3}$ & $C_{1,2}$ & \\
  \end{array}
  \right).
\end{equation*}
Here in each row the same copula type must be used.
For deriving the joint likelihood the pair-copulas have to be evaluated in conjunction with the conditional distribution functions.

Using this matrix notation \shortciteN{DissmannBrechmannCzadoKurowicka2012} give algorithms to compute the log-likelihood of an R-vine and to sample from an R-vine.
Thus maximum likelihood estimation of copula parameters is feasible.
Copula types are often selected sequentially starting from the first tree (see also \shortciteN{BrechmannCzadoAas2012}).


\section{The copula autoregressive model}\label{sect:copar}

Let $\{X_t\}_{t=1,...,T}$ and $\{Y_t\}_{t=1,...,T}$ be two univariate time series jointly observed at time point $t=1,...,T$.
The aim of this paper is to derive a flexible multivariate distribution of $\{X_t\}$ and $\{Y_t\}$, in particular allowing for non-linear dependence---serial as well as between-series.
On the one hand, such a distribution can be used to investigate the dependence among the time series (in-sample fit).
This involves for instance testing for Granger causality which will be discussed below.
On the other hand and most importantly, future values can be predicted based on this distribution (out-of-sample prediction).

Our model is based on a particular R-vine structure and defined as follows.

\begin{Definition}[COPAR]\label{def:rvar}

The copula autoregressive model (COPAR) for time series $\{X_t\}_{t=1,...,T}$ and $\{Y_t\}_{t=1,...,T}$ has the following components.
\begin{enumerate}

\item Unconditional marginal distributions $F_X$ and $F_Y$ of $\{X_t\}$ and $\{Y_t\}$, respectively.

\item An R-vine for the serial and between-series dependence of $\{X_t\}$ and $\{Y_t\}$, where the following pairs are selected.

\begin{enumerate}

\item Serial dependence of $\{X_t\}$: The pairs of a serial D-vine for $X_1,...,X_T$, i.e.
\begin{equation}
X_s,X_t|X_{s+1},...,X_{t-1},\quad 1\leq s<t\leq T.
\label{eq:rvar1}
\end{equation}

\item Between-series dependence:
\begin{equation}
X_s,Y_t|X_{s+1},...,X_t,\quad 1\leq s\leq t\leq T,
\label{eq:rvar2}
\end{equation}
and
\begin{equation}
Y_s,X_t|X_1,...,X_{t-1},Y_{s+1},...,Y_{t-1},\quad 1\leq s<t\leq T.
\label{eq:rvar3}
\end{equation}

\item Conditional serial dependence of $\{Y_t\}$: The pairs of a serial D-vine for $Y_1,...,Y_T$ conditioned on all previous values of $\{X_t\}$, i.e.
\begin{equation}
Y_s,Y_t|X_1,...,X_t,Y_{s+1},...,Y_{t-1},\quad 1\leq s<t\leq T.
\label{eq:rvar4}
\end{equation}

\end{enumerate}

Pair-copulas of the same lag length $t-s,\ t\geq s,$ are identical.
We associate 
\begin{enumerate}

\item copula $C_{t-s}^X:=C_{X_s,X_t|\boldsymbol{X}_{(s+1):(t-1)}}$ with Expression \eqref{eq:rvar1}, 

\item copulas $C_{t-s}^{XY}:=C_{X_s,Y_t|\boldsymbol{X}_{(s+1):t}}$ and $C_{t-s}^{YX}:=C_{Y_s,X_t|\boldsymbol{X}_{1:(t-1)},\boldsymbol{Y}_{(s+1):(t-1)}}$ with Expressions \eqref{eq:rvar2} and \eqref{eq:rvar3}, respectively, and

\item copula $C_{t-s}^Y:=C_{Y_s,Y_t|\boldsymbol{X}_{1:t},\boldsymbol{Y}_{(s+1):(t-1)}}$ with Expression \eqref{eq:rvar4}.

\end{enumerate}

\end{enumerate}

\end{Definition}

\begin{Remark}\label{rem:numcop}

The number of different pair-copulas utilized in the COPAR model of Definition \ref{def:rvar} is $4T-3$:
For $C_{t-s}^X$, $C_{t-s}^{YX}$ and $C_{t-s}^Y$ with $s<t$ there are $T-1$ different ones each.
In addition, there are $T$ different copulas $C_{t-s}^{XY}$, since $s\leq t$.

\end{Remark}

The joint density of the COPAR model can be derived through Expression \eqref{eq:rvinedens}.
To illustrate the rather technical definition, we present a small-dimensional example.

\begin{Example}\label{ex:dim4}

Let $T=4$.
Then the COPAR model of Definition \ref{def:rvar} for the variables $X_1,...,X_4$ and $Y_1,...,Y_4$ is constructed as shown in Figure \ref{fig:dim4}.\hfill$\square$

\begin{figure}[!t]
\centering
\includegraphics[width=.9\linewidth]{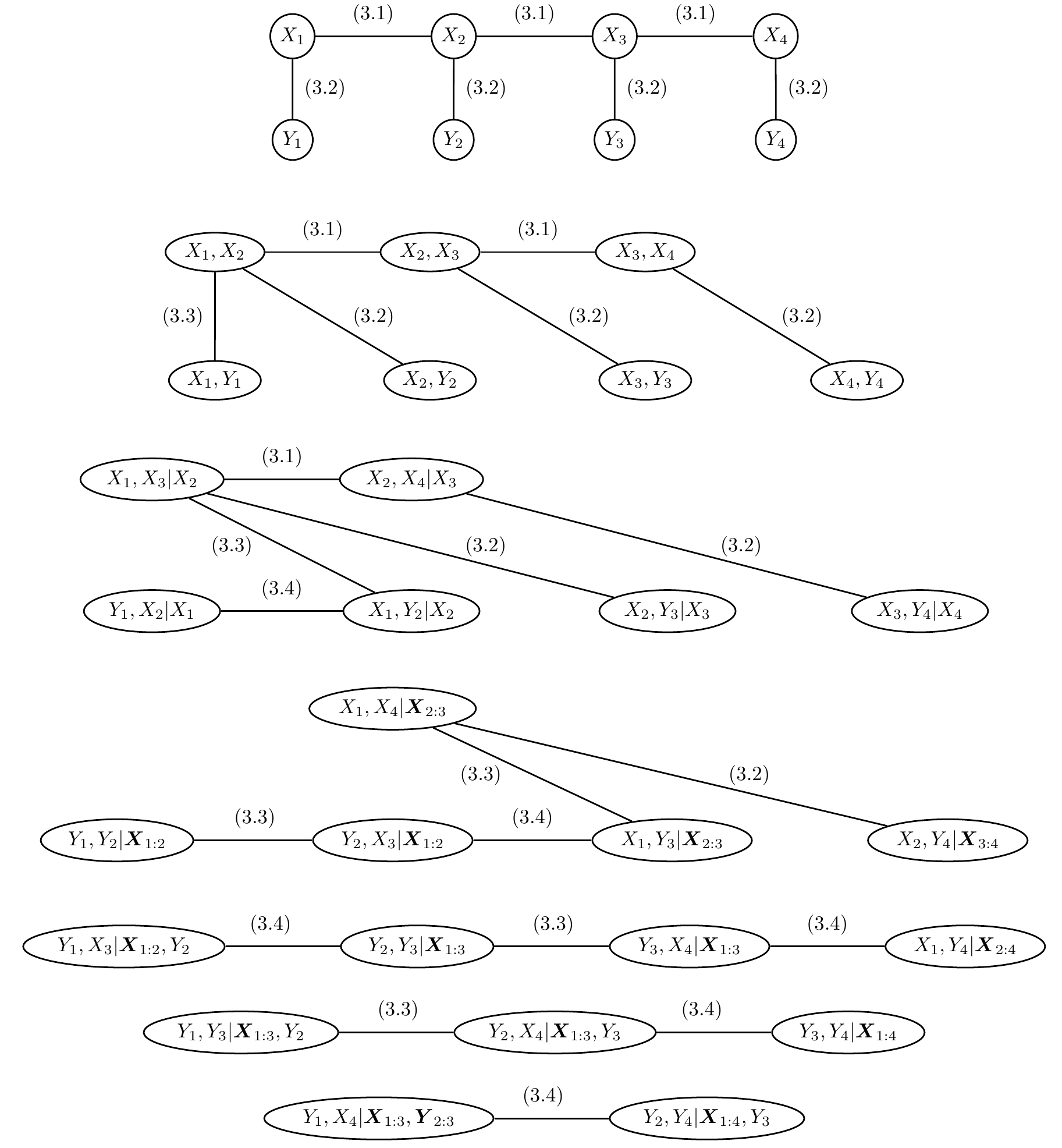}
\caption{Example for $T=4$. Edge labels relate to Definition \ref{def:rvar}.}
\label{fig:dim4}
\end{figure}

\end{Example}

Clearly, $\{X_t\}$ plays a pivotal role in this modeling approach: While the serial dependence of $\{X_t\}$ is modeled unconditionally, that of $\{Y_t\}$ is specified conditionally on $\{X_t\}$.
In other words, the roles of $\{X_t\}$ and $\{Y_t\}$ cannot be simply interchanged, since the time series play different roles.
On the other hand, the R-vine structure specifies the full joint distribution of $\{X_t\}$ and $\{Y_t\}$ so that this is mainly an issue of interpretability.
Its implications will be treated in more detail when prediction is discussed (see Section \ref{sect:pred}).
However note that the fitted joint distribution may in the end actually look differently depending on the order of variables in the modeling, since different copulas can be used.
In the case of Gaussian pair-copulas, this is not the case, since then the R-vine copula corresponds to a multivariate Gaussian copula, where the correlation matrix can be computed from conditional correlations as given by the R-vine copula parameters.

While the serial dependence is rather straightforward to understand from this model, the modeling of the between-series dependence warrants a more detailed examination.
For this purpose it is useful to look at the R-vine matrices associated to the R-vine structure of Definition \ref{def:rvar}.
Here, we continue with Example \ref{ex:dim4} first.

\begin{Example}\label{ex:dim4a}

The R-vine matrix corresponding to the R-vine structure in Figure \ref{fig:dim4} can be derived as
\begin{equation}
  \left(
  \begin{array}{cccccccc}
    \cellcolor{gray!10}Y_4 & & & \\
    Y_1 & \cellcolor{gray!10}X_4 & & \\
    Y_2 & Y_1 & \cellcolor{gray!10}Y_3 & \\
    Y_3 & Y_2 & Y_1 & \cellcolor{gray!10}X_3\\
    \cline{1-1}
    \multicolumn{1}{c|}{X_1} & Y_3 & Y_2 & Y_1 & \cellcolor{gray!10}Y_2\\
    \cline{2-3}
    X_2 & X_1 & \multicolumn{1}{c|}{X_1} & Y_2 & Y_1 & \cellcolor{gray!10}X_2\\
    \cline{4-5}
    X_3 & X_2 & X_2 & X_1 & \multicolumn{1}{c|}{X_1} & Y_1 & \cellcolor{gray!10}Y_1\\
    \cline{6-7}
    X_4 & X_3 & X_3 & X_2 & X_2 & X_1 & \multicolumn{1}{c|}{X_1} & \cellcolor{gray!10}X_1\\
  \end{array}
  \right).
\label{eq:rvmstructure4}
\end{equation}
The solid line is drawn to highlight the structure.
On the diagonal, values of $\{X_t\}$ and $\{Y_t\}$ appear alternately starting with $X_1$ and increasing from right to left.
It is clear that in this way the matrix can easily be extended to include new observations as additional columns on the left.
This will prove useful for forecasting as discussed in Section \ref{sect:pred}. 
To gain detailed insight into the structure and the dependence properties the model implies, we take all odd numbered and all even numbered columns, that is columns 1, 3, 5 and 7 and 2, 4, 6 and 8, respectively, and look at them separately.
Note however that these sub-matrices are not valid R-vine matrices themselves.
\begin{itemize}

\item \textit{Even numbered columns:}
\begin{equation}
  \left(
  \begin{array}{cccc}
  	\\
    \cellcolor{gray!10}X_4 & & \\
    Y_1 & \\
    Y_2 & \cellcolor{gray!10}X_3\\
    Y_3 & Y_1\\
    \cline{1-1}
    \multicolumn{1}{c|}{X_1} & Y_2 & \cellcolor{gray!10}X_2\\
    \cline{2-2}
    X_2 & \multicolumn{1}{c|}{X_1} & Y_1\\
    \cline{3-3}
    X_3 & X_2 & \multicolumn{1}{c|}{X_1} & \cellcolor{gray!10}X_1\\
  \end{array}
  \right).
\label{eq:structmat1}
\end{equation}
Now it becomes clear that the pairs below the solid line specify the serial dependence of $\{X_t\}$ (see Expression \eqref{eq:rvar1}).
The pairs above the solid line however model between-series dependence: Given past values of $\{X_t\}$ (and of $\{Y_t\}$), dependence of $\{X_t\}$ with regard to previous values of $\{Y_t\}$ is modeled (see Expression \eqref{eq:rvar3}).
The first column gives the following pairs: 
$Y_3,X_4|X_{1:3}$, $Y_2,X_4|X_{1:3},Y_3$ and $Y_1,X_4|X_{1:3},Y_{2:3}$.

\item \textit{Odd numbered columns:}
\begin{equation}
  \left(
  \begin{array}{cccc}
    \cellcolor{gray!10}Y_4 & & & \\
    Y_1 & & \\
    Y_2 & \cellcolor{gray!10}Y_3 & \\
    Y_3 & Y_1 \\
    \cline{1-1}
    \multicolumn{1}{c|}{X_1} & Y_2 & \cellcolor{gray!10}Y_2\\
    \cline{2-2}
    X_2 & \multicolumn{1}{c|}{X_1} & Y_1 \\
    \cline{3-3}
    X_3 & X_2 & \multicolumn{1}{c|}{X_1} & \cellcolor{gray!10}Y_1\\
    \cline{4-4}
    X_4 & X_3 & X_2 & X_1\\
  \end{array}
  \right).
\label{eq:structmat2}  
\end{equation}
Here, the pairs below the solid line also model between-series dependence, namely that of $\{Y_t\}$ with respect to previous values of $\{X_t\}$ given realizations of $\{X_t\}$ between the two time indices under consideration (see Expression \eqref{eq:rvar2}).
For example the first column specifies the following pairs: $X_4,Y_4$ (unconditional dependence), $X_3,Y_4|X_4$, $X_2,Y_4|X_{3:4}$ and $X_1,Y_4|X_{2:4}$.
Finally the pairs above the solid line specify the serial dependence structure of $\{Y_t\}$ conditioned on all observed values of $\{X_t\}$ up to the maximal time index of the pair (see Expression \eqref{eq:rvar4}).
If $\{X_t\}$ and $\{Y_t\}$ are independent, the dependence structure of $\{Y_t\}$ also is an unconditional serial D-vine.

\end{itemize}
The R-vine copula matrices corresponding to these two sub-matrices are then given as follows:
\begin{itemize}

\item \textit{Even numbered columns:}
\begin{equation*}
  \left(
  \begin{array}{cccC{.8cm}}
  	\\
    \\
    C_{Y_1X_4|\boldsymbol{X}_{1:3}\boldsymbol{Y}_{2:3}}\\
    C_{Y_2X_4|\boldsymbol{X}_{1:3}Y_3}\\
    C_{Y_3X_4|\boldsymbol{X}_{1:3}} & C_{Y_1X_3|\boldsymbol{X}_{1:2}Y_2}\\
    \cline{1-1}
    \multicolumn{1}{c|}{C_{X_1X_4|\boldsymbol{X}_{2:3}}} & C_{Y_2X_3|\boldsymbol{X}_{1:2}}\\
    \cline{2-2}
    C_{X_2X_4|X_3} & \multicolumn{1}{c|}{C_{X_1X_3|X_2}} & C_{Y_1X_2|X_1}\\
    \cline{3-3}
    C_{X_3X_4} & C_{X_2X_3} & \multicolumn{1}{c|}{C_{X_1X_2}} & \\
  \end{array}
  \right)
  =
  \left(
  \begin{array}{cccC{.3cm}}
  	\\
    \\
    C_3^{YX}\\
    C_2^{YX}\\
    C_1^{YX} & C_2^{YX}\\
    \cline{1-1}
    \multicolumn{1}{c|}{C_3^X} & C_1^{YX}\\
    \cline{2-2}
    C_2^X & \multicolumn{1}{c|}{C_2^X} & C_1^{YX}\\
    \cline{3-3}
    C_1^X & C_1^X & \multicolumn{1}{c|}{C_1^X} & \\
  \end{array}
  \right).
\end{equation*}

\item \textit{Odd numbered columns:}
\begin{equation*}
  \left(
  \begin{array}{cccc}
  	\\
    \\
    C_{Y_1Y_4|\boldsymbol{X}_{1:4}\boldsymbol{Y}_{2:3}}\\
    C_{Y_2Y_4|\boldsymbol{X}_{1:4}Y_3}\\
    C_{Y_3Y_4|\boldsymbol{X}_{1:4}} & C_{Y_1Y_3|\boldsymbol{X}_{1:3}Y_2}\\
    \cline{1-1}
    \multicolumn{1}{c|}{C_{X_1Y_4|\boldsymbol{X}_{2:4}}} & C_{Y_2Y_3|\boldsymbol{X}_{1:3}}\\
    \cline{2-2}
    C_{X_2Y_4|\boldsymbol{X}_{3:4}} & \multicolumn{1}{c|}{C_{X_1Y_3|\boldsymbol{X}_{2:3}}} & C_{Y_1Y_2|\boldsymbol{X}_{1:2}}\\
    \cline{3-3}
    C_{X_3Y_4|X_4} & C_{X_2Y_3|X_3} & \multicolumn{1}{c|}{C_{X_1Y_2|X_2}}\\
    \cline{4-4}
    C_{X_4Y_4} & C_{X_3Y_3} & C_{X_2Y_2} & C_{X_1Y_1} \\
  \end{array}
  \right)
  =
  \left(
  \begin{array}{cccc}
  	\\
    \\
    C_3^Y\\
    C_2^Y\\
    C_1^Y & C_2^Y\\
    \cline{1-1}
    \multicolumn{1}{c|}{C_3^{XY}} & C_1^Y\\
    \cline{2-2}
    C_2^{XY} & \multicolumn{1}{c|}{C_2^{XY}} & C_1^Y\\
    \cline{3-3}
    C_1^{XY} & C_1^{XY} & \multicolumn{1}{c|}{C_1^{XY}}\\
    \cline{4-4}
    C_0^{XY} & C_0^{XY} & C_0^{XY} & C_0^{XY} \\
  \end{array}
  \right).
\end{equation*}
 
\end{itemize}
This illustrates the notation of pair-copulas in Definition \ref{def:rvar}.
For the serial D-vine dependence of $\{X_t\}$ and $\{Y_t\}$ it is clear that copulas for the same lag length are identical.
For instance $C_1^X= C_{X_1X_2}= C_{X_2X_3}= C_{X_3X_4}$.
This translates to the other copulas: Copulas with the same lag length of the conditioned nodes are identical, e.g. $C_1^{YX}= C_{Y_1X_2|X_1}= C_{Y_2X_3|\boldsymbol{X}_{1:2}}= C_{Y_3X_4|\boldsymbol{X}_{1:3}}$.
When combining these two matrices, one obtains the complete R-vine copula matrix corresponding to the structure matrix \eqref{eq:rvmstructure4} utilizing 13 different pair-copulas.\hfill$\square$

\end{Example}

\begin{figure}
\centering
\includegraphics[width=.4\linewidth]{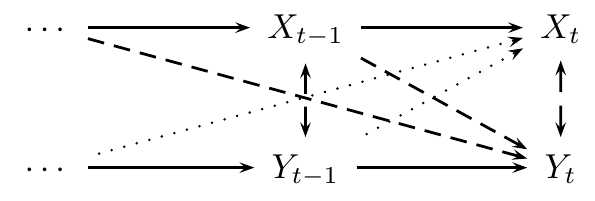}
\caption{Simplified illustration of the dependence structure of the COPAR model. Solid lines represent serial dependence, dashed and dotted lines between-series dependence given by the entries specified below the solid line in matrix \eqref{eq:structmat2} and above the solid line in matrix \eqref{eq:structmat1}, respectively.}
\label{fig:timedep}
\end{figure}

Similar to Matrix \eqref{eq:rvmstructure4} the general R-vine structure matrix of the joint distribution of $\{X_t\}_{t=1,...,T}$ and $\{Y_t\}_{t=1,...,T}$ as defined in Definition \ref{def:rvar} can be derived as
\begin{equation}
  \left(
  \begin{array}{C{0.8cm}C{0.8cm}C{0.8cm}C{0.8cm}C{0.7cm}C{0.7cm}C{0.7cm}C{0.7cm}C{0.7cm}C{0.7cm}C{0.7cm}C{0.7cm}}
    \cellcolor{gray!10}$Y_T$ & & & \\
    $Y_1$ & \cellcolor{gray!10}$X_T$ & & \\
    $Y_2$ & $Y_1$ & \cellcolor{gray!10}$Y_{T-1}$ & \\
    $\vdots$ & $Y_2$ & $Y_1$ & \cellcolor{gray!10}$X_{T-1}$ & \\
    $\vdots$ & $\vdots$ & $Y_2$ & $Y_1$ & $\ddots$\\
    $Y_{T-1}$ & $\vdots$ & $\vdots$ & $Y_2$ & $\ddots$ & $\ddots$\\
    \cline{1-1}
    \multicolumn{1}{c|}{X_1} & $Y_{T-1}$ & $Y_{T-2}$ & $\vdots$ & $\ddots$ & $\ddots$ & $\ddots$\\
    \cline{2-3}
    $X_2$ & $X_1$ & \multicolumn{1}{c|}{X_1} & $Y_{T-2}$ & & $\ddots$ & $\ddots$ & \cellcolor{gray!10}$X_3$\\
    \cline{4-4}
    $\vdots$ & $X_2$ & $X_2$ & $X_1$ & & & $\ddots$ & $Y_1$ & \cellcolor{gray!10}$Y_2$\\
    $\vdots$ & $\vdots$ & $\vdots$ & $X_2$ & & & \multicolumn{1}{c|}{} & $Y_2$ & $Y_1$ & \cellcolor{gray!10}$X_2$\\
    \cline{8-9}
    $\vdots$ & $\vdots$ & $\vdots$ & $\vdots$ & & & & $X_1$ & \multicolumn{1}{c|}{X_1} & $Y_1$ & \cellcolor{gray!10}$Y_1$ \\ 
    \cline{10-11}
    $X_T$ & $X_{T-1}$ & $X_{T-1}$ & $X_{T-2}$ & $\cdots$ & $\cdots$ & $\cdots$ & $X_2$ & $X_2$ & $X_1$ & \multicolumn{1}{c|}{X_1} & \cellcolor{gray!10}$X_1$\\       
  \end{array}
  \right).
\label{eq:rvmstructure}
\end{equation}
The interpretation of this $2T$-by-$2T$-matrix is the same as in the above example.
The corresponding copula matrix is also found in exactly the same way and therefore not shown here.
Figure \ref{fig:timedep} depicts the dependence structure specified by the different blocks of the matrices.

As noted in Remark \ref{rem:numcop} the COPAR model requires $4T-3$ copulas for dependence modeling, while a standard R-vine specification would use $T(2T-1)$. 
In other words, the number of copulas grows only linearly in the number of time points in contrast to quadratic growth of a standard R-vine.
This number even reduces when assuming an autoregressive structure of a specific order.
We therefore define the following model.

\begin{Definition}[COPAR($k$)]\label{def:copar}

The copula autoregressive model of order $k$ (COPAR($k$)) is defined as the COPAR model of Definition \ref{def:rvar}, where all pair-copulas corresponding to a lag length greater than $k$ are independence copulas, that is
\begin{equation*}
C_{t-s}^X = C_{t-s}^{XY} = C_{t-s}^{YX} = C_{t-s}^Y = \Pi \quad\text{for}\quad t-s>k,
\end{equation*}
where $\Pi$ denotes the independence copula.

\end{Definition}

\begin{Remark}\label{rem:numcop2}

Let $k$ denote the autoregressive order, then the number of different pair-copulas in the COPAR($k$) model of Definition \ref{def:copar} is $4k+1$.

\end{Remark}

Clearly, the COPAR model of Definition \ref{def:rvar} then corresponds to a COPAR($T-1$) model.
We continue Examples \ref{ex:dim4} and \ref{ex:dim4a} to illustrate this definition.

\begin{Example}\label{ex:dim4b}

If we assume a COPAR(1) model for the dependence of $X_1,...,X_4$ and $Y_1,...,Y_4$, some of the copulas specified in the R-vine copula matrix of Example \ref{ex:dim4} simplify to independence copulas.
For example $C_2^Y$ and $C_3^{XY}$ are independence copulas.
The sub-matrices that we considered before can therefore be reduced as follows, where independence copulas are indicated by the symbol $\Pi$.
The left matrix corresponds to even numbered columns, the right to odd numbered columns:
\begin{equation*}
  \left(
  \begin{array}{cccC{0.3cm}}
  	\\
    \\
    \Pi\\
    \Pi\\
    C_1^{YX} & \Pi\\
    \cline{1-1}
    \multicolumn{1}{c|}{\Pi} & C_1^{YX}\\
    \cline{2-2}
    \Pi & \multicolumn{1}{c|}{\Pi} & C_1^{YX}\\
    \cline{3-3}
    C_1^X & C_1^X & \multicolumn{1}{c|}{C_1^X} & \\
  \end{array}
  \right),\quad
    \left(
  \begin{array}{cccc}
  	\\
    \\
    \Pi\\
    \Pi\\
    C_1^Y & \Pi\\
    \cline{1-1}
    \multicolumn{1}{c|}{\Pi} & C_1^Y\\
    \cline{2-2}
    \Pi & \multicolumn{1}{c|}{\Pi} & C_1^Y\\
    \cline{3-3}
    C_1^{XY} & C_1^{XY} & \multicolumn{1}{c|}{C_1^{XY}}\\
    \cline{4-4}
    C_0^{XY} & C_0^{XY} & C_0^{XY} & C_0^{XY} \\
  \end{array}
  \right).
\end{equation*} 
The number of different pair-copulas thus reduces from 13 to 5.
Similarly, assuming an autoregressive structure of order 2 would still reduce the number of copulas to 9.\hfill$\square$

\end{Example}

Clearly, the number of different pair-copulas (see Remark \ref{rem:numcop2}) no longer depends on the number of time points $T$ and thus allows for a very parsimonious modeling.
For example, if an autoregressive structure of order 2 is assumed, only 9 different copulas are required for describing the dependence of $\{X_t\}_{t=1,...,T}$ and $\{Y_t\}_{t=1,...,T}$, while $T$ may be 1000 or greater.
In terms of model parameters this means that only a small number is needed, since most common bivariate copulas have at most two parameters and also marginal distributions have rarely more than three or four parameters.
In particular, compared to the model complexity of the bivariate VAR model \eqref{eq:var} there is essentially no difference: While a VAR($k$) model requires $4k+1$ parameters ($k$ 2-by-2-matrices with 4 entries each as well as the off-diagonal entry $\Sigma_{12}$ of the covariance matrix $\Sigma$) for modeling the serial and the between-series dependence, the COPAR($k$) model needs exactly the same number if only one parameter copulas are used.
Additionally, a VAR model uses two constant regressors as well as two residual variances (the diagonal entries $\Sigma_{11}$ and $\Sigma_{22}$ of the covariance matrix $\Sigma$) to model the margins.
In contrast, a COPAR model needs the full specification of marginal distributions, which however usually does not require many parameters (e.g. for skew-normal margins as used in Section \ref{sect:applic} below, six parameters are needed for the COPAR model in contrast to four needed by the VAR model).


\subsection{Extension to higher dimensions}

The above construction for two time series $\{X_t\}_{t=1,...,T}$ and $\{Y_t\}_{t=1,...,T}$ can also be extended to an arbitrary number of time series.
The following example illustrates the case of three time series $\{X_t\}$, $\{Y_t\}$ and $\{Z_t\}$.

\begin{Example}

Again let $T=4$.
Similarly to Definition \ref{def:rvar} and Examples \ref{ex:dim4} and \ref{ex:dim4a} an appropriate R-vine structure for the variables $X_1,...,X_4,Y_1,...,Y_4,Z_1,...,Z_4$ is constructed as the following R-vine structure matrix:
\begin{equation}
  \left(
  \begin{array}{cccccccccccc}
    \cellcolor{gray!10}Z_4 \\
    Z_1 & \cellcolor{gray!10}Y_4\\
    Z_2 & Z_1 & \cellcolor{gray!10}X_4\\
    Z_3 & Z_2 & Z_1 & \cellcolor{gray!10}Z_3\\
    \cline{1-1}
    \multicolumn{1}{c|}{Y_1} & Z_3 & Z_2 & Z_1 & \cellcolor{gray!10}Y_3\\
    \cline{2-2}
    Y_2 & \multicolumn{1}{c|}{Y_1} & Z_3 & Z_2 & Z_1 & \cellcolor{gray!10}X_3\\
    \cline{3-4}
    Y_3 & Y_2 & Y_1 & \multicolumn{1}{c|}{Y_1} & Z_2 & Z_1 & \cellcolor{gray!10}Z_2\\
    \cline{5-5}
    Y_4 & Y_3 & Y_2 & Y_2 & \multicolumn{1}{c|}{Y_1} & Z_2 & Z_1 & \cellcolor{gray!10}Y_2\\
    \cline{1-2}\cline{6-7}
    X_1 & \multicolumn{1}{c|}{X_1} & Y_3 & Y_3 & Y_2 & Y_1 & \multicolumn{1}{c|}{Y_1} & Z_1 & \cellcolor{gray!10}X_2\\
    \cline{3-5}\cline{8-8}
    X_2 & X_2 & X_1 & X_1 & \multicolumn{1}{c|}{X_1} & Y_2 & Y_2 & \multicolumn{1}{c|}{Y_1} & Z_1 & \cellcolor{gray!10}Z_1\\
    \cline{6-8}\cline{9-10}
    X_3 & X_3 & X_2 & X_2 & X_2 & X_1 & X_1 & \multicolumn{1}{c|}{X_1} & Y_1 & \multicolumn{1}{c|}{Y_1} & \cellcolor{gray!10}Y_1\\
    \cline{9-11}
    X_4 & X_4 & X_3 & X_3 & X_3 & X_2 & X_2 & X_2 & X_1 & X_1 & \multicolumn{1}{c|}{X_1} & \cellcolor{gray!10}X_1\\
  \end{array}
  \right).
\label{eq:rvmstructured3}
\end{equation}


The patterns clearly resemble those of the two-dimensional case.
Serial dependence of $\{X_t\}$ and $\{Y_t\}$ is again modeled using serial D-vine structures (for $\{Y_t\}$ conditionally on observed values of $\{X_t\}$).
In the same way, the serial dependence of $\{Z_t\}$ is captured by a serial D-vine structure conditioned on observed values of $\{X_t\}$ and $\{Y_t\}$, that means in terms of the pairs $Z_s,Z_t|X_1,...,X_t,Y_1,...,Y_t,Z_{s+1},...,Z_{t-1},\ 1\leq s<t\leq 4$.
Between-series dependence of $\{Y_t\}$ and $\{Z_t\}$ is also specified conditionally on values of $\{X_t\}$.\hfill$\square$

\end{Example}

Along the lines of this example, multivariate time series can be modeled by iteratively conditioned D-vines and appropriate between-series copulas.
Let $m\in\mathbb{N}$ be the number of different time series $\{X_{tj}\},\ j=1,...,m$, where $\{X_{t1}\}$ is the pivotal time series, $\{X_{t2}\}$ the second pivot, and so on.
Using the notation of Definition \ref{def:rvar}, an adequate R-vine based autoregressive model can then be specified in terms of $m$ blocks of $T-1$ pair-copulas $C_{t-s}^{X_j}$ each for (conditional) serial dependence of $\{X_{tj}\}$, $j=1,...,m,$ as well as $\binom{m}{2}$ blocks of $T$ pair-copulas $C_{t-s}^{X_iX_j}$ each for between-series dependence of $\{X_{ti}\}$ and $\{X_{tj}\}$ with $i<j$ and $\binom{m}{2}$ blocks of $T-1$ pair-copulas $C_{t-s}^{X_jX_i}$ each for between-series dependence when $i>j$.
The number of pair-copulas used in such a model is $m(T-1)+\binom{m}{2}T+\binom{m}{2}(T-1)=m^2T-m(m+1)/2$ and can again be significantly reduced by assuming an appropriate autoregressive order $k$, namely to $mk+\binom{m}{2}(k+1)+\binom{m}{2}k=m^2k+m(m-1)/2$, which is also the number of parameters used in a VAR($k$) model for between-series dependence of $m$ time series (not counting parameters used for marginal modeling).

This modeling approach opens up new possibilities in constructing flexible autoregressive models for arbitrary numbers of time series.
For simplicity and for illustrative reasons we concentrate here on the case of two time series.

\subsection{Model estimation and selection}\label{sect:est}

In this section we discuss several issues related to selection and estimation of COPAR models. 
We begin with a note on the marginal distributions.

The problem about the selection of appropriate marginal distributions is that no i.i.d. observations are available based on which the distribution could be chosen.
The selection is therefore subject to uncertainty and should be handled carefully.
We hence advocate using rather more complex versions of common univariate distributions, such as skew-normal or skew-t distributions, to be able to capture features of the data which may be otherwise disguised through the serial dependence.

With respect to parameter estimation, all parameters of a COPAR($k$) model can be estimated jointly by maximum likelihood techniques, since the model is rather parsimonious.
A common alternative in the literature is estimation using inference functions for margins (IFM) by \citeN{JoeXu1996}.
This means that first the parameters of the marginal distributions are estimated and then, given these parameters, the copula parameters.
As noted above, there is no i.i.d. data available for the margins, so that IFM estimation is not possible in our case.
Given a good selection of the marginal distributions, IFM-type estimation ignoring the serial dependence is however useful for selection of copula types in a pre-parameter-estimation step.

When constructing a COPAR($k$) model, $4k+1$ different copulas need to be chosen, where $k$ denotes the autoregressive order.
Since for example the copula $C_{X_1X_3|X_2}$ depends on the copulas $C_{X_1X_2}$ and $C_{X_2X_3}$ through its input arguments, copula selection will proceed sequentially.
Figure \ref{fig:selection} depicts the interdependencies of the copulas for a COPAR(2) model, more details are given in Appendix \ref{sect:supp}.
Copula selection itself can be done for example using the AIC to penalize copula families with more parameters.
Note that models can also be estimated in this sequential way.
Resulting IFM-type estimates are typically good starting values for a full maximum likelihood estimation.

\begin{figure}
\centering
\includegraphics[width=.6\linewidth]{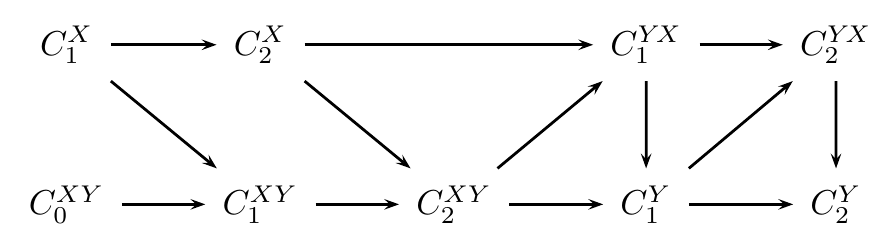}
\caption{Interdependencies of pair-copulas in a COPAR(2) model.}
\label{fig:selection}
\end{figure}

Finally, the autoregressive order has to be selected.
We propose two methods.
First, when selecting copulas, an independence test can be used to check whether the independence copula is appropriate.
If all copulas with lag greater than $k^\ast$ are selected as independence, then $k^\ast$ is the selected autoregressive order of the COPAR model.
Alternatively, one may fit different COPAR($k$) models for different lag lengths $k\geq 1$.
The optimal lag $k^\ast$ can then be chosen such that the COPAR($k^\ast$) model minimizes an information criterion such as the Akaike (AIC), Bayesian (BIC) or Hannan-Quinn (HQC):
\begin{equation}
\begin{split}
\text{AIC}(k) & = -2\hat{\ell}_k + 2p_k,\\
\text{BIC}(k) & = -2\hat{\ell}_k + \log(2T)p_k,\\
\text{HQC}(k) & = -2\hat{\ell}_k + 2\log(\log(2T))p_k,
\end{split}
\label{eq:infocrit}
\end{equation}
where $\hat{\ell}_k$ denotes the estimated log likelihood of the COPAR($k$) model and $p_k$ the number of its parameters ($4k+1$ plus marginal parameters).


\subsection{Forecasting}\label{sect:pred}

A major purpose of autoregressive modeling is forecasting.
The autoregressive R-vine model can easily be used for this.
Given time series $\{X_t\}_{t=1,...,T}$ and $\{Y_t\}_{t=1,...,T}$, we like to forecast $X_{T+h}$ and $Y_{T+h}$, where $h\geq 1$ ($h$-step prediction).

In the case $h=1$ (one-step prediction) this can be established iteratively using the following decomposition of the distribution function of $X_{T+1},Y_{T+1}|X_1,...,X_T,Y_1,...,Y_T$:
\begin{equation*}
F(x_{T+1},y_{T+1}|x_1,...,x_T,y_1,...,y_T) = F(x_{T+1}|x_1,...,x_T,y_1,...,y_T)F(y_{T+1}|x_1,...,x_{T+1},y_1,...,y_T),
\end{equation*}
where the two univariate conditional distribution functions can be described in closed form using pair-copulas of our R-vine model.
By conditional inverse sampling first of $X_{T+1}|X_1,...,X_T,Y_1,...,Y_T$ and then of $Y_{T+1}|X_1,...,X_{T+1},Y_1,...,Y_T$, a forecast can be derived.
This means that the selection which time series corresponds to $\{X_t\}$ and $\{Y_t\}$, respectively, determines which variable can be directly predicted and which conditionally.

If $h>1$, then $X_{T+h}$ and $Y_{T+h}$ can be predicted in essentially the same way by iteration: first predict $X_{T+1}$ and $Y_{T+1}$, then $X_{T+2}$ and $Y_{T+2}$ and so on.

An illustrative example provides more details.

\begin{Example}

In the setting of Example \ref{ex:dim4} we would like to predict $X_5$ and $Y_5$ given $X_1,...,X_4$ and $Y_1,...,Y_4$.
Since the dependence model of the latter eight variables is already known, the additional variables $X_5$ and $Y_5$ have to be integrated into this model appropriately such that we are able to determine the conditional distribution of $X_5,Y_5|X_1,...,X_4,Y_1,...,Y_4$.
This is straightforward using the model building principles of Definition \ref{def:rvar} and illustrated in Figure \ref{fig:pred2} in Appendix \ref{sect:addfigtab}.

In terms of the R-vine structure matrix this means the addition of two new columns:
\begin{equation*}
  \left(
  \begin{array}{cc:cccccccc}
    \cellcolor{gray!10}Y_5 & & & \\
    Y_1 & \cellcolor{gray!10}X_5\\
    Y_2 & Y_1 & \cellcolor{gray!10}Y_4\\
    Y_3 & Y_2 & Y_1 & \cellcolor{gray!10}X_4 & & \\
    Y_4 & Y_3 & Y_2 & Y_1 & \cellcolor{gray!10}Y_3 & \\
    \cline{1-1}
    \multicolumn{1}{c|}{X_1} & Y_4 & Y_3 & Y_2 & Y_1 & \cellcolor{gray!10}X_3\\
    \cline{2-3}
    X_2 & X_1 & \multicolumn{1}{c|}{X_1} & Y_3 & Y_2 & Y_1 & \cellcolor{gray!10}Y_2\\
    \cline{4-5}
    X_3 & X_2 & X_2 & X_1 & \multicolumn{1}{c|}{X_1} & Y_2 & Y_1 & \cellcolor{gray!10}X_2\\
    \cline{6-7}
    X_4 & X_3 & X_3 & X_2 & X_2 & X_1 & \multicolumn{1}{c|}{X_1} & Y_1 & \cellcolor{gray!10}Y_1\\
    \cline{8-9}
    X_5 & X_4 & X_4 & X_3 & X_3 & X_2 & X_2 & X_1 & \multicolumn{1}{c|}{X_1} & \cellcolor{gray!10}X_1\\
  \end{array}
  \right).
\end{equation*}
Using Expression \ref{eq:hfunc} the conditional distribution of $X_5|X_1,...,X_4,Y_1,...,Y_4$ can then be derived from this structure matrix as
\begin{equation*}
F(x_5|x_1,...,x_4,y_1,...,y_4) =\frac{\partial C_{Y_1X_5|\boldsymbol{X}_{1:4}\boldsymbol{Y}_{2:4}}(F(x_5|x_1,...,x_4,y_2,...,y_4),F(y_1|x_1,...,x_4,y_2,...,y_4))}{\partial F(y_1|x_1,...,x_4,y_2,...,y_4)},
\end{equation*}
where the arguments can be iteratively decomposed in the same way in terms of bivariate copulas specified in the model.
For instance,
\begin{equation*}
F(y_1|x_1,...,x_4,y_2,...,y_4) =\frac{\partial C_{Y_1Y_4|\boldsymbol{X}_{1:4}\boldsymbol{Y}_{2:3}}(F(y_1|x_1,...,x_4,y_2,y_3),F(y_4|x_1,...,x_4,y_2,y_3))}{\partial F(y_4|x_1,...,x_4,y_2,y_3)},
\end{equation*}
and so on.
However note that the model requires copulas $C_4^X=C_{X_1X_5|\boldsymbol{X}_{1:4}}$ and $C_4^{YX}=C_{Y_1X_5|\boldsymbol{X}_{1:4}\boldsymbol{Y}_{2:4}}$.
These copulas however are not known from the model and have to be taken as independence copulas.
In general, this will never be a problem, since in a COPAR($k$) model with $k\leq T-1$ (here $T=4$) these pair-copulas are independence copulas anyway and a forecast based on only four values is not quite sensible but only used for illustration here.

The case of the distribution $Y_5|X_1,...,X_5,Y_1,...,Y_4$ is similar and therefore not discussed here in more detail.\hfill$\square$

\end{Example}

Clearly, forecasting $X_{T+h}$ and $Y_{T+h}$ given $X_1,...,X_T,Y_1,...,Y_T$ simply requires to extend the R-vine matrices according to the model construction principles and then determining the conditional distribution functions in terms of bivariate conditional distribution functions which are derived from the corresponding bivariate copulas.
An appropriate autoregressive structure of order $k$ ensures that all required copulas are well-defined and also simplifies the computational burden significantly, since not all observed values have to be taken into account in the modeling.

Finally a note on conditional forecasting.
In order to forecast $Y_{T+1}$ given $X_1,...,X_T,Y_1,...,Y_T$, we derived the conditional distribution of $Y_{T+1}|X_1,...,X_{T+1},Y_1,...,Y_T$, which depends on the value of $X_{T+1}$.
To obtain a joint forecast of $X_{T+1}$ and $Y_{T+1}$, the variable $X_{T+1}$ is also predicted.
If however the true (observed) value of $X_{T+1}$ is employed here, this is called conditional forecasting, which is a very attractive tool of autoregressive models.
On the one hand, it is interesting in economics, since, for example, economic indicators such as inflation and unemployment rates are often not released simultaneously.
Let us assume that inflation rates are released first.
This information could then be used to obtain more accurate forecasts of unemployment rates.
Additionally, conditional forecasting is very useful for scenario analysis, for instance to investigate the impact of shocks to markets.


\subsection{Granger causality}

Granger causality of a time series $\{Y_t\}_{t=1,...,T}$ with respect to another time series $\{X_t\}_{t=1,...,T}$ means that $\{Y_t\}$ provides statistically significant information about $\{X_t\}$, in other words, it is helpful to predict $\{X_t\}$.
This is often denoted as $\{Y_t\}\to\{X_t\}$ (``$\{Y_t\}$ Granger causes $\{X_t\}$.'').
On the other hand, $\{Y_t\}_{t=1,...,T}$ does not Granger cause $\{X_t\}_{t=1,...,T}$ if it does not have any explanatory power with respect to future observations $X_{T+s},\ s\geq 1$.

Granger causality of one time series on another can easily be investigated using the COPAR model.
As discussed above, the COPAR model directly gives the conditional distribution of $X_{T+1}|X_1,...,X_T,Y_1,...,Y_T$.
If $\{Y_t\}$ does not Granger cause $\{X_t\}$, then this distribution is equal to that of $X_{T+1}|X_1,...,X_T$, which means that all the pairs specifying between-series dependence of $\{X_t\}$ and $\{Y_t\}$ (see Expressions \eqref{eq:rvar2} and \eqref{eq:rvar3}) are independent.
The model then collapses to two independent serial D-vines for $\{X_t\}$ and $\{Y_t\}$, respectively.
When pair-copulas of these two serial D-vines are chosen as for the pairs in the full model, then the two models are nested (most copulas, in particular those that will be considered in this paper, include the independence copula as special or boundary case) and a standard likelihood-ratio test can be used to investigate Granger causality.
Let $\ell_{RV}$ denote the log likelihood of the joint model of $\{X_t\}$ and $\{Y_t\}$ and $\ell_{DV}$ the cumulative one of the two separate D-vines for $\{X_t\}$ and $\{Y_t\}$ constructed with the same pair-copulas as the joint model, then
\begin{equation}
2(\ell_{RV}-\ell_{DV}) \stackrel{as.}{\sim} \chi_{p_{RV}-p_{DV}}^2,
\label{eq:grangerlrt}
\end{equation}
where $\chi_q^2$ denotes a $\chi^2$ distribution with $q$ degrees of freedom and $p_{RV}$ and $p_{DV}$ denote the number of parameters of the full (R-vine) and the reduced (two D-vines) model, respectively.
If copula families are however chosen differently, then a test for non-nested hypotheses such as the one by \citeN{Vuong1989} can be used.

To summarize, in order to investigate Granger causality of a time series $\{Y_t\}_{t=1,...,T}$ on another time series $\{X_t\}_{t=1,...,T}$, the COPAR model can be used in conjunction with a likelihood-ratio test.
If Granger causality of $\{X_t\}$ on $\{Y_t\}$ is to be investigated, the roles of the two time series have to be interchanged. 


\section{Applications}\label{sect:applic}

In this section we discuss three relevant applications.
First, we analyze four monthly macro-economic indicators pairwisely, namely inflation and interest rates as well as stock returns and industrial production.
Since this is the classical area of application of VAR models it is particularly interesting to see what COPAR models can add here.
Second, COPAR models are used to model daily electricity load demands in four Australian states.
Due to geography there is a strong interdependence among states which needs to be captured.
Finally, monthly Fama bond portfolio returns with medium duration are analyzed.

In each of the three data sets we compare the COPAR model to relevant benchmark models in terms of out-of-sample predictive ability:
First, VAR($k$) models are fitted using the \textit{R}-package \texttt{vars} \cite{Pfaff2008}.
Second, we also fit standard copula models with AR($k$) and AR($k$)-GARCH(1,1) margins, where the distribution of the innovations is chosen as in the corresponding COPAR model and the copula used is selected according to the AIC from a range of different copulas capturing all types of dependence (tail-independent Gaussian (N) and Frank (F), symmetric-tail-dependent Student-t, lower-tail-dependent Clayton (C), upper-tail-dependent Gumbel (G) and Joe (J), and their survival versions (SC, SG, SJ)).
Note that the copula model with AR($k$)-GARCH(1,1) is a very tough competitor model, since it allows for time-varying variances, which the COPAR and the VAR models do not.
Since the use of copula-GARCH models is a common tool for the analysis of financial time series, it is also included in the analysis.

For COPAR models we distinguish
\begin{itemize}

\item unconditional prediction of $X_{T+1}$ given $X_1,...,X_T,Y_1,...,Y_T$,

\item joint prediction of $X_{T+1}$ and $Y_{T+1}$ given $X_1,...,X_T,Y_1,...,Y_T$, and

\item conditional prediction of $Y_{T+1}$ given $X_1,...,X_{T+1},Y_1,...,Y_T$.

\end{itemize}
To speed up computations IFM-type sequential estimation as described in Section \ref{sect:est} is used for out-of-sample prediction, since the model is re-estimated for each additional prediction.
In all three applications model parameters estimated in this way were very close to full maximum likelihood estimates. 
Copulas are selected as described in Section \ref{sect:est} and from the same list as above.

Predictions are evaluated in terms of two different loss functions.
On the one hand, we consider the classical root mean squared error
\begin{equation*}
\text{RMSE}(\hat{\boldsymbol{x}};\boldsymbol{x}) = \sqrt{\frac{1}{T^\ast}\sum_{t=1}^{T^\ast} \left(\hat{x}_i-x_i\right)^2},
\end{equation*}
where $T^\ast$ denotes the number of out-of-sample predictions and $\hat{\boldsymbol{x}}=(\hat{x}_1,...,\hat{x}_{T^\ast})^\prime$ are point forecasts of $\boldsymbol{x}=(x_1,...,x_{T^\ast})^\prime$.
On the other hand, in order to check the coverage of empirical prediction intervals, mean interval scores by \citeN{GneitingRaftery2007} are taken into account:
\begin{equation*}
\text{MIS}_\alpha(\hat{\boldsymbol{l}},\hat{\boldsymbol{u}};\boldsymbol{x}) = \frac{1}{T^\ast}\sum_{t=1}^{T^\ast} \left[\big(\hat{u}_i-\hat{l}_i\big)+\frac{2}{\alpha}\big(\hat{l}_i-x_i\big)\boldsymbol{1}_{\{x_i<\hat{l}_i\}}+\frac{2}{\alpha}\big(x_i-\hat{u}_i\big)\boldsymbol{1}_{\{\hat{u}_i<x_i\}}\right],
\end{equation*}
where $\hat{\boldsymbol{l}}=(\hat{l}_1,...,\hat{l}_{T^\ast})^\prime$ and $\hat{\boldsymbol{u}}=(\hat{u}_1,...,\hat{u}_{T^\ast})^\prime$, and $[\hat{l}_i,\hat{u}_i],\ i=1,...,T^\ast,$ are $100(1-\alpha)\%$ prediction intervals.
For the copula-based models these are determined as $(\alpha/2)$ and $(1-\alpha/2)$ sample quantiles; for the VAR model closed form expressions using normal quantiles are used.


\subsection{Macro-economic indicators}

The first application investigates the interdependencies of four monthly US macro-economic indicators:
continuously compounded real returns on the S\&P500 index (\texttt{MKT}),
real interest rates of 30-day US Treasury Bills (\texttt{RRF}),
continuously compounded growth rates of the US CPI (\texttt{INF}) and
continuously compounded growth rates of US industrial production (\texttt{IPG}).
The data is available at http://www.economagic.com/ and covers 493 monthly observations from December 1959 to December 2000, which we split into a training set of 393 and a testing set of 100 observations.
Analyses of comparable data can be found in \citeN{Lee1992} and in \citeN[Chapter 11]{ZivotWang2006}.

\begin{figure}[t]
\centering
\includegraphics[width=\linewidth]{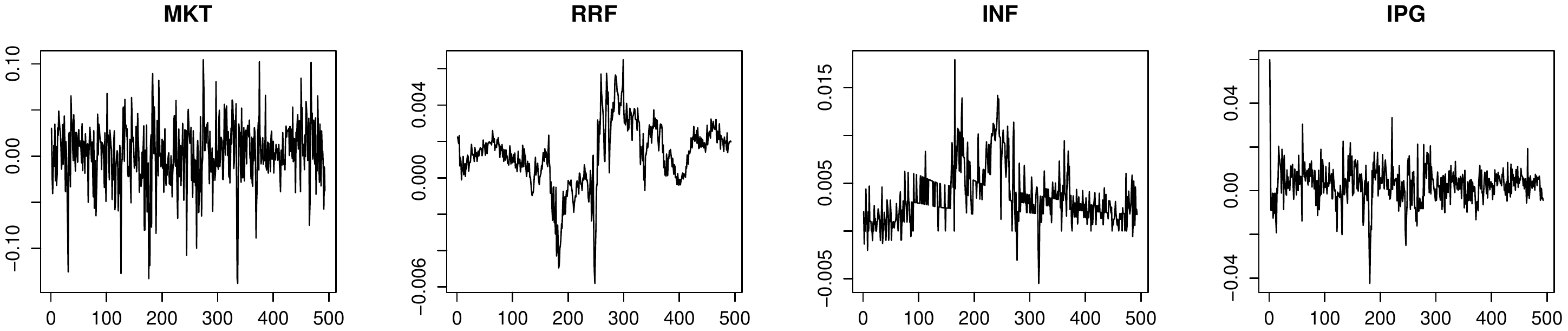}
\caption{Macro-economic indicators: Time series.}
\label{fig:econ}
\end{figure}

The time series are shown in Figure \ref{fig:econ}.
From the very different behavior of the series one may expect only little dependence between them.
A look at rank correlations (not corrected for serial dependence) confirms this, since most empirical values are below 0.1 in absolute terms and only between inflation rates and stock returns as well as inflation rates and interest rates there is a somewhat stronger negative dependence of $-0.21$ and $-0.26$, respectively.
Therefore, it will be particularly interesting to analyze Granger causality for this data.

For the marginal time series we choose skew-normal distributions to account for skewness in the data, a common feature of many economic time series.
In particular, we use the parametrization by \citeN{Azzalini1985}, where a shape parameter controls the skewness.
If the shape parameter is 0, there is no skewness and the distribution reduces to the normal one.


In the following, we investigate the causal relationships of all twelve (ordered) pairs of these four variables to see how the indicators influence each other.

\begin{figure}[t]
\centering
\includegraphics[width=.75\linewidth]{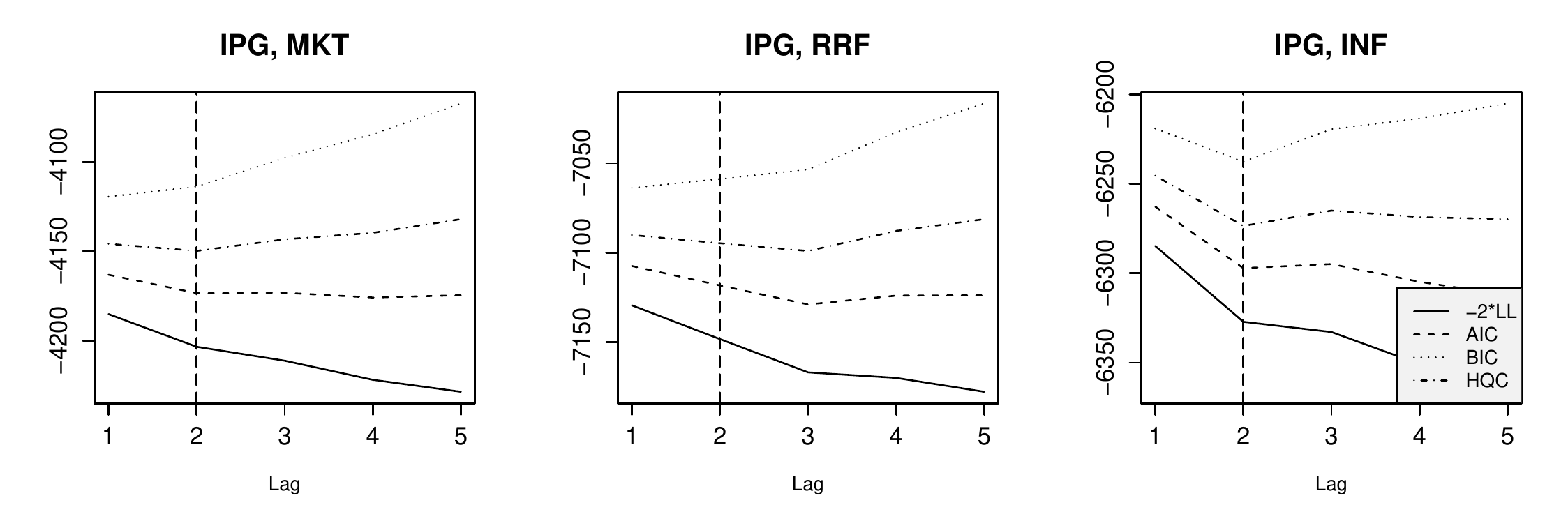}
\caption{Macro-economic indicators: Information criteria \eqref{eq:infocrit} for the fit of three variable pairs at different lag lengths, where all pair-copulas are chosen as Gaussian.}
\label{fig:econ_lag2}
\end{figure}

We select the lag length as described in Section \ref{sect:est}.
Three of the twelve diagnostic plots are shown in Figure \ref{fig:econ_lag2}.
To facilitate estimation, all pair-copulas were chosen as Gaussian; Gaussian pair-copulas are chosen as proxy, since they can model positive as well as negative symmetric dependence and only use one parameter.
The plots indicate an appropriate autoregressive order of $k=2$.

We then fitted a COPAR(2) model for the training data as described in the introduction to this section.
Selected copula types and estimated parameters are shown in Table \ref{tab:econ_par} in Appendix \ref{sect:addfigtab}. 
Although many tail-symmetric copulas (Gaussian, Student-t, Frank) were chosen, this dependence is mostly non-Gaussian and there is also evidence for some asymmetric dependence, e.g. the serial dependence of inflation rates.
AIC, BIC and HQC \eqref{eq:infocrit} confirmed this: Using non-Gaussian copulas clearly improved the in-sample fit over an only Gaussian COPAR(2) model.
In addition, there is also clear indication of skewness in the data as determined by the estimated marginal parameters.

Next we compare unconditional and joint prediction using the COPAR(2) model to predictions obtained from VAR(2), copula-AR(2) and copula-AR(2)-GARCH(1,1) models on the testing set of 100 out-of-sample observations.
In addition, we investigate whether predictions can be improved through conditional prediction, when the future realization of one variable is already known.

Tables \ref{tab:econ_pred} and \ref{tab:econ_pred2} show root mean squared errors and mean interval scores for all models and all twelve variable pairs based on 100'000 samples.
While the COPAR(2) model almost exclusively outperforms the VAR(2) model, the inclusion of GARCH(1,1) effects in the standard copula-AR(2) model still provides better predictions.
Nevertheless, the standard copula-AR(2) model is inferior to the COPAR(2) model which indicates that serial dependence is not linear unlike modeled by the AR-margins.
Rather surprisingly, conditional prediction is not useful for this data.
This however makes sense in light of the small dependence between series described above.

\begin{table}
\begin{center}
\begin{tabular}{l|rrr|rrr}
& \texttt{MKT}$|$\texttt{RRF} & \texttt{MKT}$|$\texttt{INF} & \texttt{MKT}$|$\texttt{IPG} & \texttt{RRF}$|$\texttt{MKT} & \texttt{RRF}$|$\texttt{INF} & \texttt{RRF}$|$\texttt{IPG} \\
\hline
uncond. COPAR(2) & 0.02908 & 0.02869 & 0.02920 & 0.00043 & 0.00041 & 0.00042 \\ 
joint COPAR(2) & 0.02924 & \textbf{0.02861} & 0.02921 & 0.00044 & \textbf{0.00040} & 0.00042 \\ 
VAR(2) & 0.02900 & 0.02867 & 0.02923 & 0.00042 & 0.00042 & 0.00043 \\
\hline 
copula-AR(2) & 0.02943 & 0.02948 & 0.02945 & 0.00046 & 0.00045 & 0.00046 \\ 
copula-AR(2)-GARCH(1,1) & \textbf{0.02897} & 0.02893 & \textbf{0.02895} & \textbf{0.00041} & 0.00041 & \textbf{0.00041} \\
\hline
cond. COPAR(2) & 0.02924 & 0.02912 & 0.02928 & 0.00043 & 0.00041 & 0.00043 \\  
\hline
\multicolumn{2}{c}{}\\
\end{tabular}
\begin{tabular}{l|rrr|rrr}
 & \texttt{INF}$|$\texttt{MKT} & \texttt{INF}$|$\texttt{RRF} & \texttt{INF}$|$\texttt{IPG} & \texttt{IPG}$|$\texttt{MKT} & \texttt{IPG}$|$\texttt{RRF} & \texttt{IPG}$|$\texttt{INF} \\
\hline
uncond. COPAR(2) & \textbf{0.00144} & \textbf{0.00141} & 0.00144 & 0.00437 & 0.00441 & 0.00443 \\ 
joint COPAR(2) & 0.00144 & 0.00143 & \textbf{0.00143} & \textbf{0.00433} & \textbf{0.00437} & 0.00443 \\ 
VAR(2) & 0.00148 & 0.00144 & 0.00147 & 0.00446 & 0.00445 & 0.00445 \\ 
\hline
copula-AR(2) & 0.00156 & 0.00156 & 0.00156 & 0.00498 & 0.00498 & 0.00498 \\ 
copula-AR(2)-GARCH(1,1) & 0.00145 & 0.00146 & 0.00146 & 0.00437 & 0.00438 & \textbf{0.00437} \\
\hline
cond. COPAR(2) & 0.00152 & 0.00141 & 0.00143 & 0.00437 & 0.00434 & 0.00444 \\
\hline
\end{tabular}
\caption{Macro-economic indicators: Root mean squared errors of unconditional, joint and conditional predictions from COPAR(2) models as well as of predictions from VAR(2) and copula-AR(2)-(GARCH(1,1)-)models for all twelve variable pairs and 100 out-of-sample values. Best performing methods (other than the conditional) are indicated in bold.}
\label{tab:econ_pred}
\end{center}
\end{table}  
\begin{table}
\begin{center}
\begin{tabular}{l|rrr|rrr}
& \texttt{MKT}$|$\texttt{RRF} & \texttt{MKT}$|$\texttt{INF} & \texttt{MKT}$|$\texttt{IPG} & \texttt{RRF}$|$\texttt{MKT} & \texttt{RRF}$|$\texttt{INF} & \texttt{RRF}$|$\texttt{IPG} \\
\hline
uncond. COPAR(2) & 0.1688 & 0.1639 & 0.1693 & 0.0024 & 0.0023 & 0.0023 \\ 
joint COPAR(2) & 0.1707 & 0.1649 & 0.1705 & 0.0026 & 0.0027 & 0.0028 \\ 
VAR(2) & 0.1649 & \textbf{0.1606} & 0.1646 & 0.0029 & 0.0029 & 0.0029 \\
\hline 
copula-AR(2) & 0.1703 & 0.1712 & 0.1706 & 0.0030 & 0.0030 & 0.0030 \\ 
copula-AR(2)-GARCH(1,1) & \textbf{0.1611} & 0.1611 & \textbf{0.1622} & \textbf{0.0021} & \textbf{0.0021} & \textbf{0.0021} \\
\hline 
cond. COPAR(2) & 0.1716 & 0.1626 & 0.1705 & 0.0026 & 0.0026 & 0.0026 \\ 
\hline
\multicolumn{2}{c}{}\\
\end{tabular}
\begin{tabular}{l|rrr|rrr}
 & \texttt{INF}$|$\texttt{MKT} & \texttt{INF}$|$\texttt{RRF} & \texttt{INF}$|$\texttt{IPG} & \texttt{IPG}$|$\texttt{MKT} & \texttt{IPG}$|$\texttt{RRF} & \texttt{IPG}$|$\texttt{INF} \\
\hline
uncond. COPAR(2) & \textbf{0.0084} & 0.0083 & \textbf{0.0084} & 0.0312 & 0.0316 & 0.0308 \\ 
joint COPAR(2) & 0.0084 & \textbf{0.0082} & 0.0084 & 0.0318 & 0.0316 & 0.0310 \\ 
VAR(2) & 0.0094 & 0.0092 & 0.0093 & 0.0317 & 0.0320 & 0.0311 \\
\hline 
copula-AR(2) & 0.0094 & 0.0093 & 0.0094 & 0.0325 & 0.0325 & 0.0325 \\ 
copula-AR(2)-GARCH(1,1) & 0.0089 & 0.0089 & 0.0089 & \textbf{0.0284} & \textbf{0.0285} & \textbf{0.0286} \\
\hline 
cond. COPAR(2) & 0.0086 & 0.0081 & 0.0084 & 0.0316 & 0.0310 & 0.0310 \\ 
\hline
\end{tabular}
\caption{Macro-economic indicators: Mean interval scores of unconditional, joint and conditional 95\% prediction intervals from COPAR(2) models as well as of 95\% prediction intervals from VAR(2) and copula-AR(2)-(GARCH(1,1)-)models for all twelve variable pairs and 100 out-of-sample values. Best performing methods (other than the conditional) are indicated in bold.}
\label{tab:econ_pred2}
\end{center}
\end{table}

\begin{figure}[t]
\centering
\includegraphics[width=.4\linewidth]{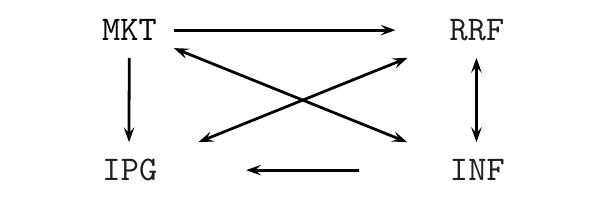}
\includegraphics[width=.4\linewidth]{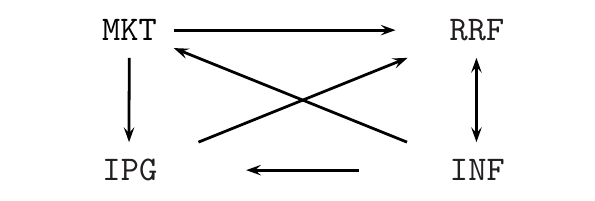}
\caption{Macro-economic indicators: Granger causality of variables according to the COPAR(2) model (left panel) and the VAR(2) model (right panel) at the 5\% level.}
\label{fig:econ_ganger}
\end{figure}

Finally, we investigate Granger causality between the time series using the likelihood ratio test \eqref{eq:grangerlrt}.
The resulting causalities as identified by COPAR(2) and VAR(2) models are indicated by arrows in Figure \ref{fig:econ_ganger}.
The two models mainly identify the same interdependencies, differences are likely due to the non-linear modeling of the COPAR model.
In particular, inflation rates are determined to significantly influence all other variables, while industrial production growth only causes interest rates.
Nonetheless, although there is Granger causality among most variable pairs, conditional predictions did not improve the forecasting accuracy.

\subsection{Australian electricity load demands}

\begin{figure}[t]
\centering
\includegraphics[width=\linewidth]{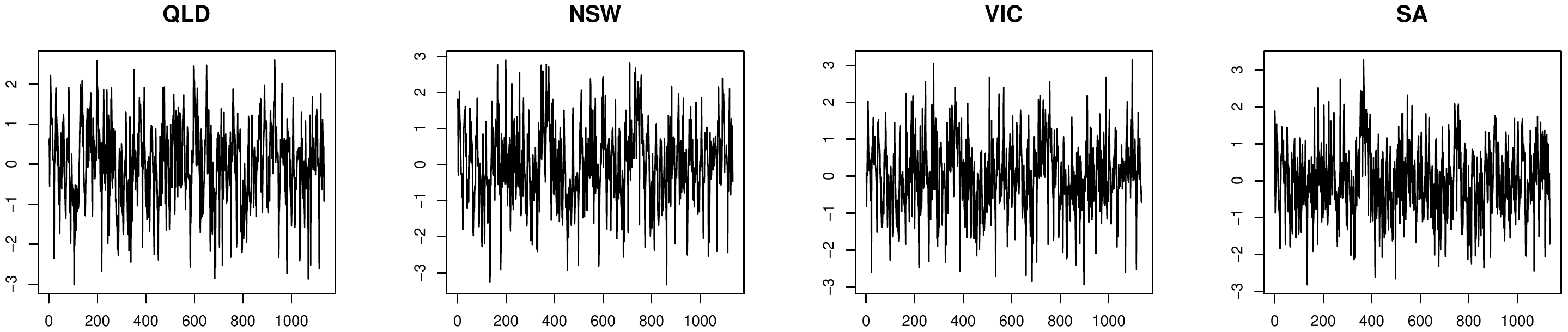}
\caption{Electricity load demands: Time series.}
\label{fig:energy}
\end{figure}

In this application we analyze average daily electricity load demands in Gigawatt for the Australian states Queensland (\texttt{QLD}), New South Wales (\texttt{NSW}), Victoria (\texttt{VIC}) and South Australia (\texttt{SA}).
The observations have been calculated by averaging the half-hourly observed data for one day which are available at http://www.aemo.com.au/.
The observed time period is from May 16, 2005 to June 30, 2008 with 1135 daily observations, split into a training set of 635 and a testing set of 500 observations.
Figure \ref{fig:energy} shows the time series after preprocessing the data to remove trend and weekly and annual seasonalities.

To allow for possible skewness in the data we again choose the skew-normal distribution by \citeN{Azzalini1985} for the margins.
In contrast to the previous application there however is clear between-series dependence.
Unconditional empirical between-series dependence (not corrected for serial dependence in the margins) is shown in Table \ref{tab:energy_dep}.
Since the Australian infrastructure is concentrated near the coast, dependence is observed in particular for the pairs \texttt{QLD}-\texttt{NSW}, \texttt{NSW}-\texttt{VIC} and \texttt{VIC}-\texttt{SA}, which mirror the geographical location of the states on the coastline.
While focusing on these pairs, we investigate the causal relationships of all twelve ordered pairs of these four states.

\begin{table}[t]
\begin{center}
\begin{tabular}{r|rrrr}
& \texttt{QLD} & \texttt{NSW} & \texttt{VIC} & \texttt{SA} \\ 
\hline
\texttt{QLD} & 1.00 & 0.40 & 0.13 & 0.08 \\ 
\texttt{NSW} & 0.40 & 1.00 & 0.49 & 0.39 \\ 
\texttt{VIC} & 0.13 & 0.49 & 1.00 & 0.68 \\ 
\texttt{SA} & 0.08 & 0.39 & 0.68 & 1.00 \\
\hline
\end{tabular}
\caption{Electricity load demands: Unconditional empirical rank correlations between time series (not corrected for serial dependence).}
\label{tab:energy_dep}
\end{center}
\end{table}

The lag length of COPAR($k$) models for daily electricity load demands is selected as $k=2$ according to different information criteria as described in Section \ref{sect:est}.
The comparison of COPAR(2) models with Gaussian copulas and with arbitrary copulas then shows that dependence is mainly tail-symmetric but mostly non-Gaussian.
In particular, tail-dependent Student-t copulas play a central role in describing the between-series dependence.
Such tail dependence is ignored by Gaussian copulas and therefore again stresses the need for non-Gaussian modeling.
Selected copula types and estimated parameters are shown in Table \ref{tab:energy_par} in Appendix \ref{sect:addfigtab}.

The out-of-sample predictive ability of the COPAR(2) model is evaluated on the testing set of 500 observations.
Evaluation criteria are computed based on 100'000 samples.
Tables \ref{tab:energy_pred} and \ref{tab:energy_pred2} show that the COPAR(2) model performs very strongly and clearly outperforms the VAR(2) model and in most cases even the copula-AR(2)-GARCH(1,1) model.

Due to the clear geographical relationships between the states, conditional prediction clearly improves the forecasting accuracy.
In particular when the demand of the neighboring state is known, there is a significant improvement.
This result indicates that the COPAR model may be useful for scenario analysis, e.g. to examine the effect of shocks like an extreme demand to the electricity market.

\begin{table}
\begin{center}
\begin{tabular}{l|rrr|rrr}
& \texttt{QLD}$|$\texttt{NSW} & \texttt{QLD}$|$\texttt{VIC} & \texttt{QLD}$|$\texttt{SA} & \texttt{NSW}$|$\texttt{QLD} & \texttt{NSW}$|$\texttt{VIC} & \texttt{NSW}$|$\texttt{SA} \\ 
\hline
uncond. COPAR(2) & \textbf{0.653} & \textbf{0.648} & \textbf{0.645} & \textbf{0.692} & \textbf{0.674} & \textbf{0.679} \\ 
joint COPAR(2) & 0.663 & 0.670 & 0.657 & 0.706 & 0.674 & 0.689 \\ 
VAR(2) & 0.674 & 0.670 & 0.668 & 0.697 & 0.690 & 0.690 \\
\hline  
copula-AR(2) & 0.673 & 0.674 & 0.673 & 0.697 & 0.697 & 0.697 \\ 
copula-AR(2)-GARCH(1,1) & 0.679 & 0.679 & 0.679 & 0.699 & 0.699 & 0.698 \\ 
\hline
cond. COPAR(2) & 0.606 & 0.624 & 0.627 & 0.654 & 0.603 & 0.639 \\
\hline
\multicolumn{2}{c}{}\\
\end{tabular}
\begin{tabular}{l|rrr|rrr}
& \texttt{VIC}$|$\texttt{QLD} & \texttt{VIC}$|$\texttt{NSW} & \texttt{VIC}$|$\texttt{SA} & \texttt{SA}$|$\texttt{QLD} & \texttt{SA}$|$\texttt{NSW} & \texttt{SA}$|$\texttt{VIC} \\ 
\hline
uncond. COPAR(2) & 0.676 & \textbf{0.674} & \textbf{0.666} & 0.691 & 0.686 & 0.691 \\ 
joint COPAR(2) & \textbf{0.672} & 0.679 & 0.684 & 0.690 & 0.686 & 0.679 \\ 
VAR(2) & 0.693 & 0.692 & 0.685 & 0.687 & \textbf{0.684} & 0.688 \\
\hline 
copula-AR(2) & 0.690 & 0.690 & 0.689 & \textbf{0.686} & 0.687 & \textbf{0.685} \\ 
copula-AR(2)-GARCH(1,1) & 0.692 & 0.692 & 0.692 & 0.687 & 0.687 & 0.687 \\
\hline
cond. COPAR(2) & 0.638 & 0.598 & 0.555 & 0.666 & 0.644 & 0.560 \\
\hline
\end{tabular}
\caption{Electricity load demands: Root mean squared errors of unconditional, joint and conditional predictions from COPAR(2) models as well as of predictions from VAR(2) and copula-AR(2)-(GARCH(1,1)-)models for all twelve variable pairs and 500 out-of-sample values. Best performing methods (other than the conditional) are indicated in bold.}
\label{tab:energy_pred}
\end{center}
\end{table}

\begin{table}
\begin{center}
\begin{tabular}{l|rrr|rrr}
& \texttt{QLD}$|$\texttt{NSW} & \texttt{QLD}$|$\texttt{VIC} & \texttt{QLD}$|$\texttt{SA} & \texttt{NSW}$|$\texttt{QLD} & \texttt{NSW}$|$\texttt{VIC} & \texttt{NSW}$|$\texttt{SA} \\ 
\hline
uncond. COPAR(2) & \textbf{3.567} & \textbf{3.551} & \textbf{3.540} & \textbf{3.771} & 3.838 & 3.836 \\ 
joint COPAR(2) & 3.701 & 3.705 & 3.612 & 3.818 & \textbf{3.790} & 3.887 \\ 
VAR(2) & 3.901 & 3.855 & 3.839 & 3.788 & 3.987 & 3.968 \\ 
\hline
copula-AR(2) & 3.881 & 3.883 & 3.865 & 3.789 & 3.797 & \textbf{3.792} \\ 
copula-AR(2)-GARCH(1,1) & 3.697 & 3.706 & 3.705 & 3.843 & 3.849 & 3.845 \\
\hline 
cond. COPAR(2) & 3.126 & 3.291 & 3.258 & 3.292 & 3.039 & 3.280 \\ 
\hline
\multicolumn{2}{c}{}\\
\end{tabular}
\begin{tabular}{l|rrr|rrr}
& \texttt{VIC}$|$\texttt{QLD} & \texttt{VIC}$|$\texttt{NSW} & \texttt{VIC}$|$\texttt{SA} & \texttt{SA}$|$\texttt{QLD} & \texttt{SA}$|$\texttt{NSW} & \texttt{SA}$|$\texttt{VIC} \\ 
\hline
uncond. COPAR(2) & \textbf{3.766} & \textbf{3.781} & 3.834 & 3.262 & 3.237 & 3.288 \\ 
joint COPAR(2) & 3.770 & 3.809 & 3.881 & 3.299 & 3.287 & \textbf{3.235} \\ 
VAR(2) & 3.806 & 3.782 & 3.847 & 3.275 & \textbf{3.236} & 3.280 \\ 
\hline
copula-AR(2) & 3.817 & 3.798 & \textbf{3.813} & \textbf{3.248} & 3.239 & 3.244 \\ 
copula-AR(2)-GARCH(1,1) & 3.822 & 3.818 & 3.822 & 3.263 & 3.244 & 3.252 \\ 
\hline
cond. COPAR(2) & 3.300 & 3.175 & 3.074 & 3.010 & 2.946 & 2.658 \\ 
\hline
\end{tabular}
\caption{Electricity load demands: Mean interval scores of unconditional, joint and conditional 95\% prediction intervals from COPAR(2) models as well as of 95\% prediction intervals from VAR(2) and copula-AR(2)-(GARCH(1,1)-)models for all tewlve variable pairs and 500 out-of-sample values. Best performing methods (other than the conditional) are indicated in bold.}
\label{tab:energy_pred2}
\end{center}
\end{table}

\subsection{Fama bond portfolio returns}

\begin{figure}[t]
\centering
\includegraphics[width=.6\linewidth]{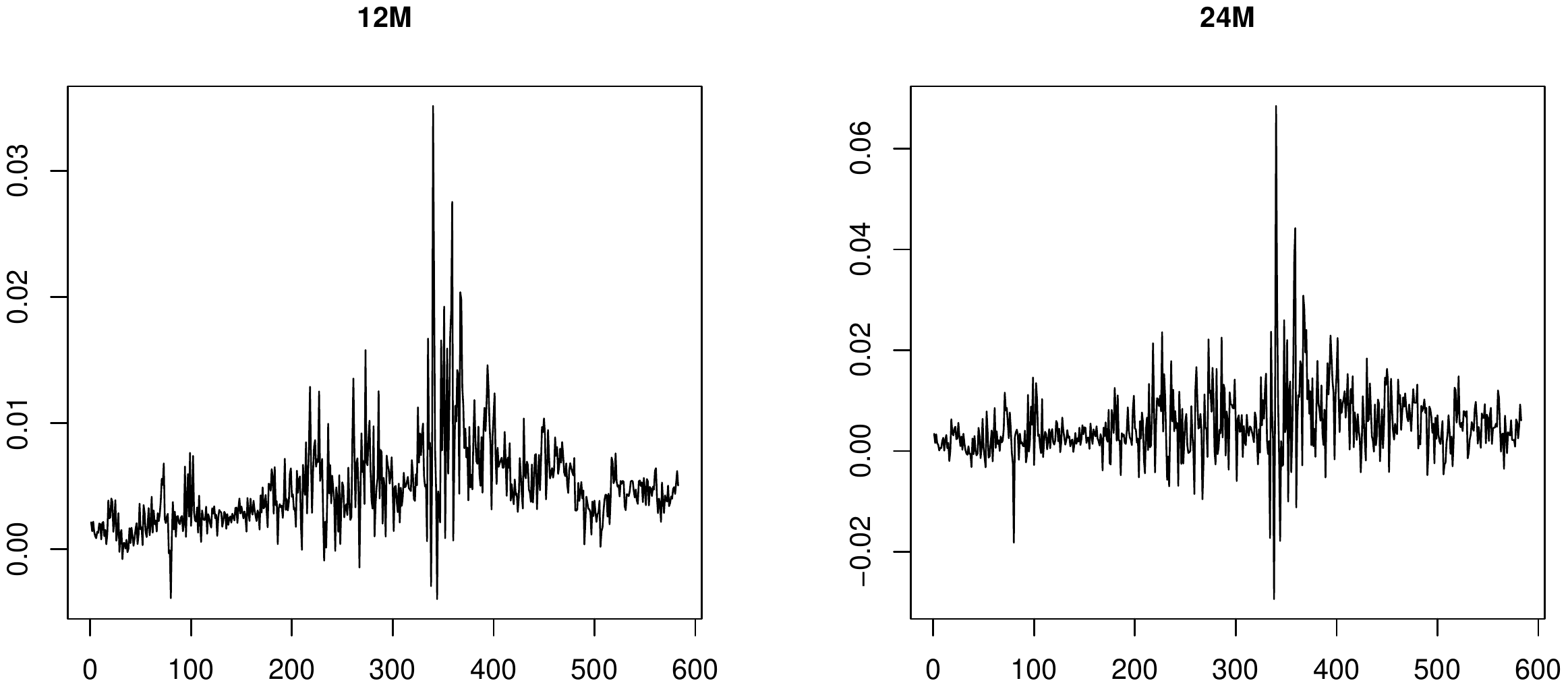}
\caption{Bond returns: Time series.
}
\label{fig:fama}
\end{figure}

After analyzing economic time series data in first two applications, we now investigate the interdependencies among financial variables, namely monthly Fama bond portfolio\footnote{Artificial zero-coupon bonds constructed after first extracting the term structure from a subset of US bonds.} returns obtained from the CRSP Center for Research in Security Prices for maturities 12 and 24 months (variables \texttt{12M} and \texttt{24M}, respectively).
The time series are observed from January 1952 to December 2003 with 684 monthly observations (training set of 584, testing set of 100 observations) and shown in Figure \ref{fig:fama}.

In contrast to the first two applications (low and medium between-series dependence, respectively), unconditional empirical between-series rank correlation (not corrected for serial dependence in the margins) between the Fama bond portfolio return series is quite strong (0.84).

Also in contrast to the first two applications we opt for a more sophisticated marginal distribution to appropriately capture the behavior of the financial time series: We choose the hyperbolic distribution which is widely used for financial return data and accounts for both fat tails and skewness (see \shortciteN{McNeilFreyEmbrechts2005}).
As for the skew-normal distribution, the  hyperbolic distribution is symmetric if the skewness parameter is 0.
An additional shape parameter controls how much weight is assigned to the tails and to the center of the distribution.

Serial dependence in the bond return data is more persistent than in the previous two applications.
Based on the discussion in Section \ref{sect:est} we select a lag length of $k=3$ and there even is some indication that maybe a higher order may be reasonable.

The in-sample fit and copula selection shows that there is clear non-Gaussian dependence in the data (see Tables \ref{tab:fama_par1} and \ref{tab:fama_par2} in Appendix \ref{sect:addfigtab}).
While Student-t and Frank copulas are central for serial and between-series dependence modeling, also tail-asymmetric Joe, Gumbel and Clayton copulas have been identified to determine the joint behavior of the time series.


Prediction evaluation criteria based on the out-of-sample testing set of 100 observations and on 100'000 samples are shown in Table \ref{tab:fama_pred}.
The COPAR(3) model's predictive ability is clearly superior to that of the VAR(3) model as well as to a standard copula-AR(3) model with and without GARCH(1,1) effects.
Also conditional prediction improves the forecasting accuracy due to the significant between-series dependence.
As before this result may be used for scenario analyses of the bond market:
How do shocks to bonds with one maturity influence bonds with the other maturity?

\begin{table}
\begin{center}
\begin{tabular}{l|rr|rr}
& \multicolumn{2}{c|}{RMSE} & \multicolumn{2}{c}{MIS} \\
& \texttt{12M}$|$\texttt{24M} & \texttt{24M}$|$\texttt{12M} & \texttt{12M}$|$\texttt{24M} & \texttt{24M}$|$\texttt{12M}\\ 
\hline
uncond. COPAR(3) & \textbf{0.00140} & 0.00431 & \textbf{0.00796} & 0.02392 \\ 
joint COPAR(3) & 0.00144 & \textbf{0.00420} & 0.00803 & \textbf{0.02110} \\
VAR(3) & 0.00165 & 0.00457 & 0.01069 & 0.02595 \\
\hline 
copula-AR(3) & 0.00171 & 0.00520 & 0.01087 & 0.02697 \\
copula-AR(3)-GARCH(1,1) & 0.00372 & 0.00472 & 0.04986 & 0.02344 \\
\hline 
cond. COPAR(3) & 0.00081 & 0.00230 & 0.00411 & 0.01207 \\
\hline
\end{tabular}
\caption{Bond returns: Root mean squared errors and mean interval scores of unconditional, joint and conditional predictions and 95\% prediction intervals from COPAR(3) models as well as of predictions and 95\% prediction intervals from VAR(3) and copula-AR(3)-(GARCH(1,1)-)models for both ordered variable pairs and 100 out-of-sample values. Best performing methods (other than the conditional) are indicated in bold.}
\label{tab:fama_pred}
\end{center}
\end{table}  


\section{Conclusion}\label{sect:concl}

In this paper we described the novel copula autoregressive model which benefits from the flexibility of R-vine copulas and allows to model non-linear dependence among multiple time series.
While three relevant applications to financial and economic time series demonstrate the usefulness of these models, their practical importance will be even more pronounced when stronger non-linear and non-symmetric dependencies are present, in time as well as between series.

The applications however also showed that time-varying variance effects may be needed in certain cases and improve over time-constant modeling.
Approaches to this issue are subject of future research.


\bibliographystyle{chicago}
\bibliography{copar-bib}


\clearpage

\appendix

\section{Technical supplement}\label{sect:supp}

As noted in Section \ref{sect:est} sequential copula selection and likelihood computation is not straightforward in the COPAR model.
Figure \ref{fig:selection} illustrates the interdependencies among the copulas.
For likelihood computation, of course the R-vine matrix specification \eqref{eq:rvmstructure} could be used instead.
However, given that the number of time points $T$ might be large, evaluation of this $2T$-by-$2T$-matrix is computationally rather inefficient, since most matrix entries do  not contain any information due to the assumed autoregressive order (see Example \ref{ex:dim4b}). 

In the following we therefore present how to sequentially select copulas in a COPAR(2) model for data $\{x_t\}_{t=1,...,T}$ and $\{y_t\}_{t=1,...,T}$.
Rather than selecting copulas, preliminarily determined copulas can be estimated or likelihoods can be evaluated by simply altering the respective lines.

\begin{enumerate}

\item Serial dependence of $\{X_t\}_{t=1,...,T}$: $X_s,X_t|X_{s+1},...,X_{t-1},\ 1\leq s<t\leq T$.

\begin{enumerate}

\item Select copula $C_1^X=C_{X_{t-1}X_t}$ based on $\{F_X(x_t)\}_{t=1,...,T-1}$ and $\{F(x_t)\}_{t=2,...,T}$.

\item Compute $F(x_t|x_{t-1}),\ t=3,...,T,$ and $F(x_{t-1}|x_t),\ t=2,...,T,$ from $C_1^X$ using Expression \eqref{eq:hfunc}. 

\item Select copula $C_2^X=C_{X_{t-2}X_t|X_{t-1}}$ based on $F(x_{t-1}|x_t)$, $t=2,...,T-1,$ and $F(x_t|x_{t-1})$, $t=3,...,T$.

\item Compute $F(x_{t-2}|x_{t-1},x_t),\ t=3,...,T,$ and $F(x_t|x_{t-2},x_{t-1}),\ t=3,...,T,$ from $C_2^X$ using Expression \eqref{eq:hfunc}. 

\end{enumerate}

\item Between-series dependence $X_s,Y_t|X_{s+1},...,X_t,\ 1\leq s\leq t\leq T$.

\begin{enumerate}

\item Select copula $C_0^{XY}=C_{X_tY_t}$ based on $\{F_X(x_t)\}_{t=1,...,T}$ and $\{F_Y(y_t)\}_{t=1,...,T}$.

\item Compute $F(y_t|x_t),\ t=2,...,T,$ from $C_0^{XY}$ using Expression \eqref{eq:hfunc}. 

\item Select copula $C_1^{XY}=C_{X_{t-1}Y_t|X_t}$ based on $F(x_{t-1}|x_t),\ t=2,...,T,$ and $F(y_t|x_t),\ t=2,...,T$.

\item Compute $F(y_t|x_{t-1},x_t),\ t=3,...,T,$ from $C_1^{XY}$ using Expression \eqref{eq:hfunc}. 

\item Select copula $C_2^{XY}=C_{X_{t-2}Y_t|\boldsymbol{X}_{(t-1):t}}$ based on $F(x_{t-2}|x_{t-1},x_t),\ t=3,...,T,$ and $F(y_t|x_{t-1},x_t),\ t=3,...,T$.

\item Compute $F(y_t|x_{t-2},x_{t-1},x_t),\ t=3,...,T-1,$ from $C_2^{XY}$ using Expression \eqref{eq:hfunc}. 

\end{enumerate}

\item Between-series dependence $Y_s,X_t|X_1,...,X_{t-1},Y_{s+1},...,Y_{t-1},\ 1\leq s<t\leq T,$ and conditional serial dependence of $\{Y_t\}_{t=1,...,T}$: $Y_s,Y_t|X_1,...,X_t,Y_{s+1},...,Y_{t-1},\ 1\leq s<t\leq T$.

\begin{enumerate}

\item Select copula $C_1^{YX}=C_{Y_{t-1}X_t|\boldsymbol{X}_{1:(t-1)}}$ based on $F(y_t|x_1,...,x_t),\ t=1,...,T-1,$ and $F(x_t|x_1,...,x_{t-1}),\ t=2,...,T$, where
\begin{equation*}
F(y_t|x_1,...,x_t) = \begin{cases} F(y_1|x_1) & t=1\\ F(y_2|x_1,x_2) & t=2\\ F(y_t|x_{t-2},x_{t-1},x_t) & t\geq 3 \end{cases},
\end{equation*}
and
\begin{equation*}
F(x_t|x_1,...,x_{t-1}) = \begin{cases} F(x_2|x_1) & t=2\\ F(x_t|x_{t-2},x_{t-1}) & t\geq 3 \end{cases}.
\end{equation*}

\item Compute $F(x_t|x_1,...,x_{t-1},y_{t-1}),\ t=3,...,T,$ and $F(y_{t-1}|x_1,...,x_t),\ t=2,...,T,$ from $C_1^{YX}$ using Expression \eqref{eq:hfunc}.

\item Select copula $C_1^Y=C_{Y_{t-1}Y_t|\boldsymbol{X}_{1:t}}$ based on $F(y_{t-1}|x_1,...,x_t),\ t=2,...,T,$ and $F(y_t|x_1,...,x_t)$, $t=2,...,T$, where
\begin{equation*}
F(y_{t-1}|x_1,...,x_t) = \begin{cases} F(y_1|x_1,x_2) & t=2\\ F(y_2|x_1,x_2,x_3) & t=3\\ F(y_{t-1}|x_{t-3},...,x_t) & t\geq 4 \end{cases}.
\end{equation*}

\item Compute $F(y_{t-1}|x_1,...,x_t,y_t),\ t=2,...,T-1,$ and $F(y_t|x_1,...,x_t,y_{t-1}),\ t=3,...,T,$ from $C_1^Y$ using Expression \eqref{eq:hfunc}. 

\item Select copula $C_2^{YX}=C_{Y_{t-2}X_t|\boldsymbol{X}_{1:(t-1)}Y_{t-1}}$ based on $F(y_{t-1}|x_1,...,x_t,y_t),\ t=2,...,T-1,$ and $F(x_t|x_1,...,x_{t-1},y_{t-1}),\ t=3,...,T$, where
\begin{equation*}
F(y_{t-1}|x_1,...,x_t,y_t) = \begin{cases} F(y_1|x_1,x_2,y_2) & t=2\\ F(y_2|x_1,x_2,x_3,y_3) & t=3\\ F(y_{t-1}|x_{t-3},x_{t-2},x_{t-1},x_t,y_t) & t\geq 4 \end{cases},
\end{equation*}
and
\begin{equation*}
F(x_t|x_1,...,x_{t-1},y_{t-1}) = \begin{cases} F(x_3|x_1,x_2,y_2) & t=3\\ F(x_4|x_1,x_2,x_3,y_3) & t=4\\ F(x_t|x_{t-4},x_{t-3},x_{t-2},x_{t-1},y_{t-1}) & t\geq 5 \end{cases}.
\end{equation*}

\item Compute $F(y_{t-2}|x_1,...,x_t,y_{t-1}),\ t=3,...,T,$ from $C_2^{YX}$ using Expression \eqref{eq:hfunc}.

\item Select copula $C_2^Y=C_{Y_{t-2}Y_t|\boldsymbol{X}_{1:t}Y_{t-1}}$ based on $F(y_{t-2}|x_1,...,x_t,y_{t-1}),\ t=3,...,T,$ and $F(y_t|x_1,...,x_t,y_{t-1})$, $t=3,...,T$, where
\begin{equation*}
F(y_{t-2}|x_1,...,x_t,y_{t-1}) = \begin{cases} F(y_1|x_1,x_2,x_3,y_2) & t=3\\ F(y_2|x_1,x_2,x_3,x_3,x_4,y_3) & t=4\\ F(y_{t-2}|x_{t-4},...,x_t,y_{t-1}) & t\geq 5 \end{cases},
\end{equation*}
and
\begin{equation*}
F(y_t|x_1,...,x_t,y_{t-1}) = \begin{cases} F(y_3|x_1,x_2,x_3,y_2) & t=3\\ F(y_4|x_1,x_2,x_3,x_4,y_3) & t=4\\ F(y_t|x_{t-4},...,x_t,y_{t-1}) & t\geq 5 \end{cases}.
\end{equation*}

\end{enumerate}

\end{enumerate}
For autoregressive orders higher than 2 the above procedure can be easily extended.


\clearpage

\section{Additional figures and tables}\label{sect:addfigtab}

\begin{figure}[!ht]
\centering
\includegraphics[width=\linewidth]{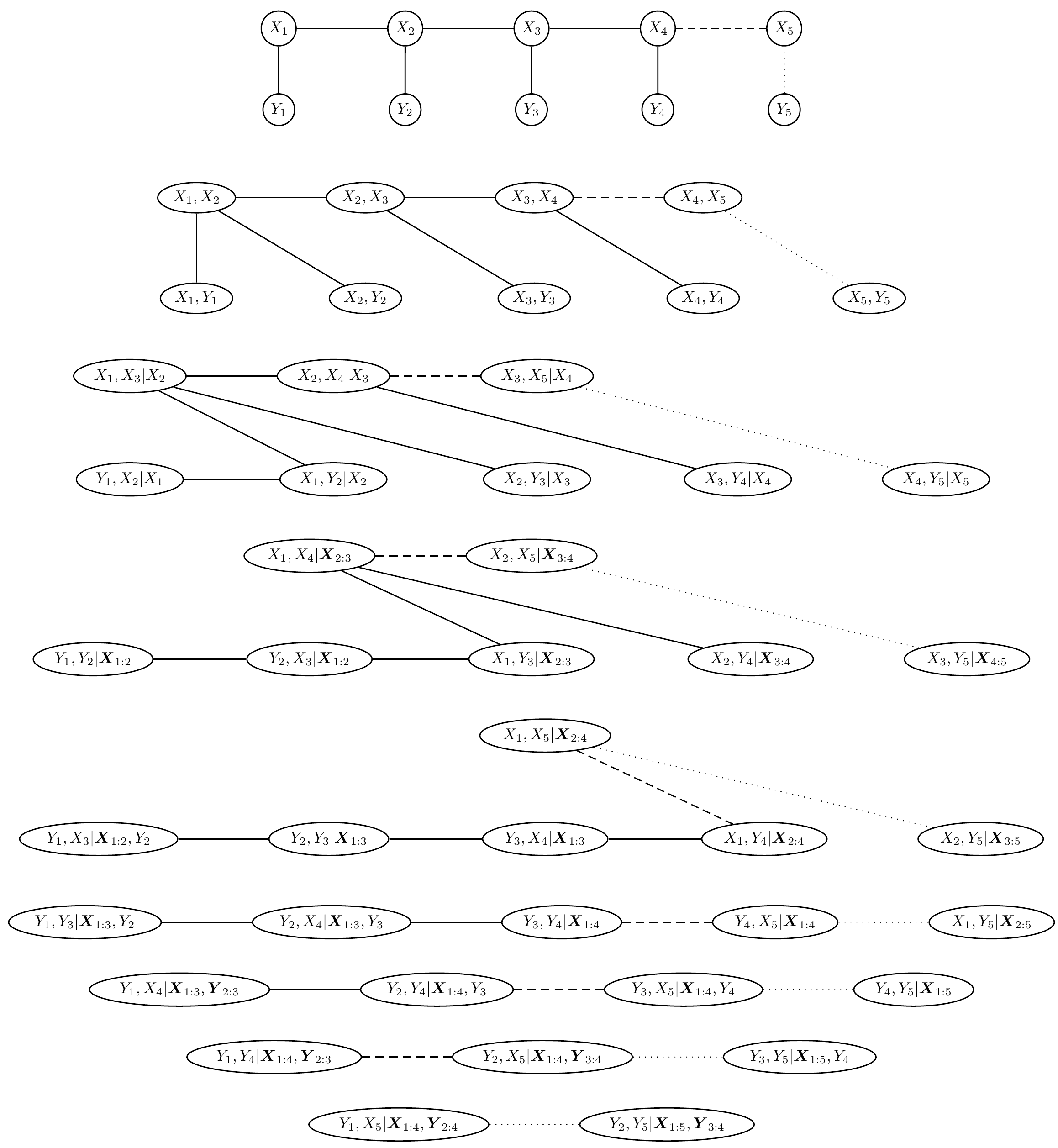}
\caption{R-vine for $X_1,...,X_4,Y_1,...,Y_4$ extended to include $X_5$ and $Y_5$. Dashed lines indicate additional edges to include $X_5$, dotted those to include $Y_5$.}
\label{fig:pred2}
\end{figure}

\begin{table}
\renewcommand{\arraystretch}{0.88}
\small
\begin{center}
\begin{tabular}{r|rrr|rr:rrr:rr:rr}
& \multicolumn{3}{c|}{Margins} & \multicolumn{9}{c}{Copulas}\\
& loc. & scale & skew & $C_1^X$ & $C_2^X$ & $C_0^{XY}$ & $C_1^{XY}$ & $C_2^{XY}$ & $C_1^{YX}$ & $C_2^{YX}$ & $C_1^Y$ & $C_2^Y$ \\ 
\hline
\texttt{MKT}, & 0.0352 & 0.0494 & -1.8383 & N & F & F & SJ & SJ & C & F & F & t \\ 
\texttt{RRF} & 0.0027 & 0.0027 & -1.3316 & 0.26 & -0.49 & -0.12 & 1.01 & 1.01 & 0.02 & -0.71 & 15.6 & 0.24 \\ 
&  &  &  &  &  &  &  &  &  &  &  & 7.17 \\ 
&  &  &  & \textit{0.17} & \textit{-0.05} & \textit{-0.01} & \textit{0.01} & \textit{0.01} & \textit{0.01} & \textit{-0.08} & \textit{0.77} & \textit{0.16} \\ 
\hline
\texttt{MKT}, & 0.0352 & 0.0494 & -1.8383 & N & F & F & N & N & N & SC & G & t \\ 
\texttt{INF} & 5e-04 & 0.0049 & 2.679 & 0.25 & -0.49 & -1.02 & -0.03 & -0.06 & -0.11 & 0.07 & 1.86 & 0.27 \\ 
&  &  &  &  &  &  &  &  &  &  &  & 9.39 \\
&  &  &  & \textit{0.16} & \textit{-0.05} & \textit{-0.11} & \textit{-0.02} & \textit{-0.04} & \textit{-0.07} & \textit{0.03} & \textit{0.46} & \textit{0.18} \\  
\hline
\texttt{MKT}, & 0.0352 & 0.0494 & -1.8383 & N & F & C & SJ & SG & N & N & t & F \\ 
\texttt{IPG} & -0.0016 & 0.0104 & 0.7242 & 0.25 & -0.52 & 0.02 & 1.01 & 1.02 & -0.07 & -0.09 & 0.40 & 1.00 \\ 
&  &  &  &  &  &  &  &  &  &  & 9.95 &  \\
&  &  &  & \textit{0.16} & \textit{-0.06} & \textit{0.01} & \textit{0.01} & \textit{0.02} & \textit{-0.04} & \textit{-0.06} & \textit{0.26} & \textit{0.11} \\ 
\hline
\texttt{RRF}, & 0.0027 & 0.0027 & -1.3316 & F & t & F & t & F & SJ & C & N & F \\ 
\texttt{MKT} &0.0352 & 0.0494 & -1.8383  & 15.52 & 0.23 & 0.21 & 0.04 & -0.41 & 1.07 & 0.13 & 0.26 & -0.41 \\ 
&  &  &  &  & 6.71 &  & 13.95 &  &  &  &  &  \\
&  &  &  & \textit{0.77} & \textit{0.15} & \textit{0.02} & \textit{0.02} & \textit{-0.05} & \textit{0.04} & \textit{0.06} & \textit{0.17} & \textit{-0.05} \\  
\hline
\texttt{RRF}, & 0.0029 & 0.0029 & -1.3316 & F & t & F & t & t & F & N & G & t \\ 
\texttt{INF} & 5e-04 & 0.005 & 2.679 & 15.42 & 0.28 & -0.58 & 0.05 & 0.07 & -1.28 & -0.04 & 1.92 & 0.28 \\ 
&  &  &  &  & 7.25 &  & 30 & 24.91 &  &  &  & 14.96 \\ 
&  &  &  & \textit{0.77} & \textit{0.18} & \textit{-0.06} & \textit{0.03} & \textit{0.04} & \textit{-0.14} & \textit{-0.03} & \textit{0.48} & \textit{0.18} \\ 
\hline
\texttt{RRF}, & 0.0028 & 0.0027 & -1.3316 & F & t & C & N & t & C & C & t & F \\ 
\texttt{IPG} & -0.0021 & 0.0104 & 0.7242 & 15.07 & 0.24 & 0.12 & -0.05 & -0.10 & 0.10 & 0.13 & 0.36 & 0.97 \\ 
&  &  &  &  & 6.48 &  &  & 10.50 &  &  & 12.24 &  \\
&  &  &  & \textit{0.76} & \textit{0.16} & \textit{0.06} & \textit{-0.03} & \textit{-0.06} & \textit{0.05} & \textit{0.06} & \textit{0.23} & \textit{0.11} \\ 
\hline
\texttt{INF}, & 5e-04 & 0.0049 & 2.679 & G & t & F & F & SC & F & N & t & F \\ 
\texttt{MKT} & 0.0352 & 0.0494 & -1.8383 & 1.91 & 0.27 & -1.62 & -0.56 & 0.09 & 0.31 & -0.05 & 0.21 & -0.53 \\ 
&  &  &  &  & 11.64 &  &  &  &  &  & 20.08 &  \\
&  &  &  & \textit{0.48} & \textit{0.17} & \textit{-0.18} & \textit{-0.06} & \textit{0.04} & \textit{0.03} & \textit{-0.03} & \textit{0.13} & \textit{-0.06} \\  
\hline
\texttt{INF}, & 5e-04 & 0.0049 & 2.679 & G & t & F & t & N & F & t & t & t \\ 
\texttt{RRF} & 0.0028 & 0.0027 & -1.3316 & 1.95 & 0.26 & -0.09 & -0.06 & -0.02 & 0.01 & -0.09 & 0.93 & 0.16 \\ 
&  &  &  &  & 12.18 &  & 30 &  &  & 17.43 & 4.61 & 4.62 \\
&  &  &  & \textit{0.49} & \textit{0.17} & \textit{-0.01} & \textit{-0.04} & \textit{-0.01} & \textit{0.00} & \textit{-0.06} & \textit{0.77} & \textit{0.10} \\  
\hline
\texttt{INF}, & 6e-04 & 0.005 & 2.679 & G & t & N & N & t & F & F & t & F \\ 
\texttt{IPG} & -0.0016 & 0.0104 & 0.7242 & 1.92 & 0.26 & -0.09 & -0.10 & -0.09 & 0.14 & -0.06 & 0.39 & 0.93 \\ 
&  &  &  &  & 12.48 &  &  & 30.00 &  &  & 9.88 &  \\
&  &  &  & \textit{0.48} & \textit{0.17} & \textit{-0.05} & \textit{-0.06} & \textit{-0.06} & \textit{0.02} & \textit{-0.01} & \textit{0.25} & \textit{0.10} \\  
\hline
\texttt{IPG}, & -0.0016 & 0.0105 & 0.7242 & t & F & C & N & N & SG & SG & N & F \\ 
\texttt{MKT} & 0.0353 & 0.0488 & -1.8383 & 0.39 & 1.03 & 0.04 & -0.09 & -0.10 & 1.03 & 1.09 & 0.25 & -0.49 \\ 
&  &  &  & 9.25 &  &  &  &  &  &  &  &  \\
&  &  &  & \textit{0.26} & \textit{0.11} & \textit{0.02} & \textit{-0.06} & \textit{-0.06} & \textit{0.02} & \textit{0.08} & \textit{0.16} & \textit{-0.05} \\  
\hline
\texttt{IPG}, & -0.0021 & 0.0104 & 0.7242 & t & F & C & SJ & SJ & F & t & F & SG \\ 
\texttt{RRF} & 0.0027 & 0.0027 & -1.3316 & 0.39 & 1.01 & 0.03 & 1.01 & 1.03 & -0.07 & -0.08 & 15.16 & 1.17 \\ 
&  &  &  & 8.48 &  &  &  &  &  & 8.88 &  &  \\
&  &  &  & \textit{0.26} & \textit{0.11} & \textit{0.02} & \textit{0.01} & \textit{0.02} & \textit{-0.01} & \textit{-0.05} & \textit{0.76} & \textit{0.14} \\ 
\hline
\texttt{IPG}, & -0.0021 & 0.0104 & 0.7242 & t & F & N & N & N & t & F & G & t \\ 
\texttt{INF} & 5e-04 & 0.005 & 2.679 & 0.41 & 0.96 & -0.09 & -0.03 & -0.03 & -0.13 & -0.71 & 1.91 & 0.26 \\ 
&  &  &  & 10.31 &  &  &  &  & 30.00 &  &  & 12.17 \\ 
&  &  &  & \textit{0.27} & \textit{0.11} & \textit{-0.06} & \textit{-0.02} & \textit{-0.02} & \textit{-0.08} & \textit{-0.08} & \textit{0.48} & \textit{0.17} \\ 
\hline
\end{tabular}
\caption{Macro-economic indicators: Estimated parameters and chosen copulas of the COPAR(2) models, where the first time series in each block corresponds to $\{X_t\}$ and the second to $\{Y_t\}$. Italic numbers indicate corresponding Kendall's $\tau$s.}
\label{tab:econ_par}
\end{center}
\end{table}

\begin{table}
\renewcommand{\arraystretch}{0.88}
\small
\begin{center}
\begin{tabular}{r|rrr|rr:rrr:rr:rr}
& \multicolumn{3}{c|}{Margins} & \multicolumn{9}{c}{Copulas}\\
& loc. & scale & skew & $C_1^X$ & $C_2^X$ & $C_0^{XY}$ & $C_1^{XY}$ & $C_2^{XY}$ & $C_1^{YX}$ & $C_2^{YX}$ & $C_1^Y$ & $C_2^Y$ \\ 
\hline
\texttt{QLD}, & 0.5347 & 1.1087 & -0.6587 & F & F & t & F & SC & SC & F & F & N \\ 
\texttt{NSW} & -0.8134 & 1.2549 & 1.0882 & 7.05 & -0.55 & 0.37 & -0.20 & 0.05 & 0.18 & -0.14 & 5.63 & -0.07 \\ 
&  &  &  &  &  & 8.63 &  &  &  &  &  &  \\
&  &  &  & \textit{0.56} & \textit{-0.06} & \textit{0.24} & \textit{-0.02} & \textit{0.02} & \textit{0.08} & \textit{-0.02} & \textit{0.49} & \textit{-0.04} \\ 
\hline 
\texttt{QLD}, & 0.5627 & 1.1083 & -0.6831 & F & F & SJ & F & F & F & N & F & N \\ 
\texttt{NSW} & -0.5783 & 1.0801 & 0.8359 & 7.03 & -0.53 & 1.14 & 0.19 & 0.06 & 0.01 & 0.10 & 5.78 & -0.12 \\ 
&  &  &  &  &  &  &  &  &  &  &  &  \\
&  &  &  & \textit{0.56} & \textit{-0.06} & \textit{0.07} & \textit{0.02} & \textit{0.01} & \textit{0.00} & \textit{0.06} & \textit{0.50} & \textit{-0.08} \\
\hline 
\texttt{QLD}, & 0.647 & 1.1597 & -0.8191 & F & F & SJ & N & F & F & F & G & F \\ 
\texttt{SA} & -0.9021 & 1.4215 & 1.9323 & 7.05 & -0.54 & 1.06 & -0.02 & -0.03 & 0.01 & 0.62 & 1.85 & -0.53 \\ 
&  &  &  &  &  &  &  &  &  &  &  &  \\
&  &  &  & \textit{0.56} & \textit{-0.06} & \textit{0.04} & \textit{-0.01} & \textit{0.00} & \textit{0.00} & \textit{0.07} & \textit{0.46} & \textit{-0.06} \\ 
\hline 
\texttt{NSW}, & -0.8824 & 1.315 & 1.1642 & F & F & t & t & G & F & G & t & F \\ 
\texttt{QLD} & 0.7313 & 1.2351 & -1.1263 & 6.46 & -0.43 & 0.39 & 0.14 & 1.05 & 0.06 & 1.04 & 0.72 & -0.52 \\ 
&  &  &  &  &  & 9.47 & 18.29 &  &  &  & 13.81 &  \\
&  &  &  & \textit{0.54} & \textit{-0.05} & \textit{0.26} & \textit{0.09} & \textit{0.05} & \textit{0.01} & \textit{0.04} & \textit{0.51} & \textit{-0.06} \\ 
\hline 
\texttt{NSW}, & -0.6892 & 1.2084 & 0.8284 & F & F & t & N & F & F & t & F & F \\ 
\texttt{VIC} & -0.6653 & 1.113 & 1.0071 & 6.36 & -0.64 & 0.44 & -0.16 & -0.03 & 2.29 & -0.10 & 4.70 & -0.58 \\ 
&  &  &  &  &  & 6.99 &  &  &  & 13.73 &  &  \\
&  &  &  & \textit{0.53} & \textit{-0.07} & \textit{0.29} & \textit{-0.1} & \textit{0.00} & \textit{0.24} & \textit{-0.07} & \textit{0.44} & \textit{-0.06} \\ 
\hline 
\texttt{NSW}, & -0.913 & 1.3292 & 1.2318 & F & F & t & F & N & N & N & t & F \\ 
\texttt{SA} & -0.6843 & 1.1507 & 1.0412 & 6.43 & -0.50 & 0.43 & -0.51 & 0.01 & 0.35 & -0.07 & 0.56 & -0.55 \\ 
&  &  &  &  &  & 5.57 &  &  &  &  & 8.65 &  \\
&  &  &  & \textit{0.54} & \textit{-0.05} & \textit{0.28} & \textit{-0.06} & \textit{0.01} & \textit{0.22} & \textit{-0.04} & \textit{0.38} & \textit{-0.06} \\ 
\hline
\texttt{VIC}, & -0.6205 & 1.0997 & 0.9191 & F & N & SJ & F & N & F & C & t & F \\ 
\texttt{QLD} & 0.7381 & 1.2083 & -1.0317 & 5.88 & -0.11 & 1.17 & -0.06 & 0.04 & 0.33 & 0.03 & 0.76 & -0.64 \\ 
&  &  &  &  &  &  &  &  &  &  & 9.18 &  \\
&  &  &  & \textit{0.51} & \textit{-0.07} & \textit{0.09} & \textit{-0.01} & \textit{0.03} & \textit{0.04} & \textit{0.02} & \textit{0.55} & \textit{-0.07} \\ 
\hline 
\texttt{VIC}, & -0.5819 & 1.0952 & 0.8255 & F & N & t & t & F & N & F & F & F \\ 
\texttt{NSW} & -0.7386 & 1.2001 & 1.0095 & 6.11 & -0.12 & 0.43 & 0.31 & 0.47 & -0.07 & 0.83 & 5.17 & 0.01 \\ 
&  &  &  &  &  & 8.66 & 8.33 &  &  &  &  &  \\
&  &  &  & \textit{0.52} & \textit{-0.07} & \textit{0.28} & \textit{0.20} & \textit{0.05} & \textit{-0.05} & \textit{0.09} & \textit{0.47} & \textit{0.00} \\
\hline 
\texttt{VIC}, & -0.6299 & 1.1162 & 0.9363 & F & N & t & N & F & t & N & G & t \\ 
\texttt{SA} & -0.7868 & 1.2349 & 1.4554 & 5.87 & -0.11 & 0.67 & -0.09 & 0.26 & 0.23 & -0.12 & 1.48 & -0.03 \\ 
&  &  &  &  &  & 24.46 &  &  & 14.38 &  &  & 14.37 \\
&  &  &  & \textit{0.51} & \textit{-0.07} & \textit{0.47} & \textit{-0.06} & \textit{0.03} & \textit{0.15} & \textit{-0.08} & \textit{0.32} & \textit{-0.02} \\ 
\hline 
\texttt{SA}, & -0.526 & 1.0738 & 0.8463 & t & N & SJ & F & F & F & SC & t & F \\ 
\texttt{QLD} & 0.7778 & 1.2763 & -1.1955 & 0.63 & -0.07 & 1.07 & 0.01 & 0.37 & 0.00 & 0.09 & 0.76 & -0.56 \\ 
&  &  &  & 10.76 &  &  &  &  &  &  & 10.11 &  \\
&  &  &  & \textit{0.43} & \textit{-0.05} & \textit{0.04} & \textit{0.00} & \textit{0.04} & \textit{0.00} & \textit{0.04} & \textit{0.55} & \textit{-0.06} \\
\hline 
\texttt{SA}, & -0.6242 & 1.1364 & 1.0308 & t & N & t & N & F & N & N & F & C \\ 
\texttt{NSW} & -0.758 & 1.2212 & 1.079 & 0.63 & -0.07 & 0.30 & 0.31 & 0.72 & -0.04 & 0.08 & 5.28 & 0.01 \\ 
&  &  &  & 10.76 &  & 5.30 &  &  &  &  &  &  \\
&  &  &  & \textit{0.44} & \textit{-0.05} & \textit{0.19} & \textit{0.20} & \textit{0.08} & \textit{-0.03} & \textit{0.05} & \textit{0.47} & \textit{0.01} \\
\hline 
\texttt{SA}, & -0.5563 & 1.0947 & 0.8318 & t & N & t & t & F & N & SC & F & N \\ 
\texttt{VIC} & -0.6184 & 1.0878 & 0.9467 & 0.64 & -0.07 & 0.65 & 0.26 & 0.21 & -0.04 & 0.05 & 4.47 & -0.04 \\ 
&  &  &  & 10.53 &  & 24.75 & 8.63 &  &  &  &  &  \\
&  &  &  & \textit{0.44} & \textit{-0.05} & \textit{0.45} & \textit{0.17} & \textit{0.02} & \textit{-0.02} & \textit{0.02} & \textit{0.42} & \textit{-0.02} \\ 
\hline
\end{tabular}
\caption{Electricity load demands: Estimated parameters and chosen copulas of the COPAR(2) models, where the first time series in each block corresponds to $\{X_t\}$ and the second to $\{Y_t\}$. Italic numbers indicate corresponding Kendall's $\tau$s.}
\label{tab:energy_par}
\end{center}
\end{table}

\begin{table}
\begin{center}
\begin{tabular}{r|rrrr}
 & shape & loc. & scale & skew \\ 
\hline
\texttt{M12}, & 0.6872 & 0.0029 & 0.0032 & 0.0021 \\ 
\texttt{M24} & 0.0368 & 0.0027 & 0.0063 & 0.0029 \\
\hline 
\texttt{M24}, & 0.0332 & 0.0024 & 0.0060 & 0.0027 \\ 
\texttt{M12} & 0.6864 & 0.0025 & 0.0027 & 0.0023 \\ 
\hline
\end{tabular}
\caption{Bond returns: Estimated marginal parameters of the COPAR(3) models.}
\label{tab:fama_par1}
\end{center}
\end{table}

\begin{table}
\renewcommand{\arraystretch}{0.88}
\small
\begin{center}
\begin{tabular}{r|rrr:rrrr:rrr:rrr}
& $C_1^X$ & $C_2^X$ & $C_3^X$ & $C_0^{XY}$ & $C_1^{XY}$ & $C_2^{XY}$ & $C_3^{XY}$ & $C_1^{YX}$ & $C_2^{YX}$ & $C_3^{YX}$ & $C_1^Y$ & $C_2^Y$ & $C_3^Y$ \\ 
\hline
\texttt{M12}, & F & t & F & G & t & F & t & t & t & F & t & F & SG \\ 
\texttt{M24} & 5.32 & 0.32 & 2.07 & 2.90 & -0.27 & -1.24 & -0.08 & -0.36 & -0.27 & -0.93 & 0.51 & 1.76 & 1.22 \\ 
&  & 5.55 &  &  & 13.96 &  & 22.13 & 10.51 & 18.47 &  & 12.96 &  &  \\ 
& \textit{0.48} & \textit{0.21} & \textit{0.22} & \textit{0.66} & \textit{-0.17} & \textit{-0.14} & \textit{-0.05} & \textit{-0.24} & \textit{-0.18} & \textit{-0.10} & \textit{0.34} & \textit{0.19} & \textit{0.18} \\ 
\hline
\texttt{M24}, & J & t & F & G & SC & t & t & SC & SC & SC & t & t & t \\ 
\texttt{M12} & 1.34 & 0.09 & 1.32 & 2.59 & 0.15 & 0.09 & 0.07 & 0.35 & 0.12 & 0.07 & 0.76 & 0.43 & 0.34 \\ 
&  & 8.09 &  &  &  & 18.07 & 17.87 &  &  &  & 2.99 & 16.98 & 3.37 \\
& \textit{0.16} & \textit{0.06} & \textit{0.14} & \textit{0.61} & \textit{0.07} & \textit{0.06} & \textit{0.04} & \textit{0.15} & \textit{0.05} & \textit{0.03} & \textit{0.55} & \textit{0.28} & \textit{0.22} \\  
\hline
\end{tabular}
\caption{Bond returns: Estimated parameters and chosen copulas of the COPAR(3) models, where the first time series in each block corresponds to $\{X_t\}$ and the second to $\{Y_t\}$. Italic numbers indicate corresponding Kendall's $\tau$s.}
\label{tab:fama_par2}
\end{center}
\end{table}

\end{document}